\documentclass[journal,twoside,web]{ieeecolor}
\usepackage{generic}
\usepackage{cite}
\usepackage{amsmath,amssymb,amsfonts}
\usepackage{amsmath,amsbsy,amssymb,bm,mathrsfs}
\usepackage{algorithmic}
\usepackage{graphicx}
\usepackage{algorithm,algorithmic}
\usepackage{hyperref}
\hypersetup{hidelinks=true}
\usepackage{textcomp}

\usepackage{bbm}

\usepackage{mathtools}
\usepackage{bbm}
\usepackage{tabularray}

\usepackage{booktabs}
\usepackage{pdflscape}
\usepackage{multirow}
\usepackage{tabstackengine}
\usepackage[labelformat=simple]{subcaption}

\def\BibTeX{{\rm B\kern-.05em{\sc i\kern-.025em b}\kern-.08em
    T\kern-.1667em\lower.7ex\hbox{E}\kern-.125emX}}
\newtheorem{lemma}{Lemma}
\newtheorem{proposition}{Proposition}
\newtheorem{theorem}{Theorem}

\newtheorem{definition}{Definition}
\newtheorem{assumption}{Assumption}

\def\5n{\negthinspace \negthinspace \negthinspace \negthinspace \negthinspace }
\def\4n{\negthinspace \negthinspace \negthinspace \negthinspace }
\def\3n{\negthinspace \negthinspace \negthinspace }
\def\2n{\negthinspace \negthinspace }

\def\proof {{Proof.} }
\def\endproof{\hfill $\Box$ \vskip .3cm}

\def \d{{\rm d}}

\def \0{\bm{0}}

\def \r{\bm{r}}
\def \s{\bm{s}}
\def \f{\bm{f}}
\def \g{\bm{g}}
\def \k{\bm{k}}

\def \0{\bm 0}
\def \m1{\mathbb{1}}

\def \mR{\mathbb{R}}
\def \mP{\mathbb{P}}
\def \mQ{\mathbb{Q}}

\def \1{\mathbbm{1}}

\def \cP{\mathcal{P}}
\def \cF{\mathcal{F}} 
\def \cA{\mathcal{A}}

\def \mv{\textrm{mv}}
\def \tg{\textrm{tg}}

\def \VSSM{\textrm{VSSM}}

\newcommand{\Var}{{\rm Var}}
\newcommand{\E}{{\rm E}}
\newcommand{\Cov}{{\rm Cov}}

\begin{document}
\title{Dynamic Factor Model-Based Multiperiod Mean-Variance Portfolio Selection with Portfolio Constraints}

\author{Jianjun Gao,\IEEEmembership{Member, IEEE}, Chengneng Jin, Yun Shi,  Xiangyu Cui
	\thanks{This research is partially supported by the National Natural Science Foundation of China under grants 71971083, 71931004, 71971132, 72171138,  72192832 and 72201067.}
	\thanks{Jianjun Gao is with the School of Information Managment and Engineering, Shanghai University of Finance and Economics, Shanghai 200433 China (e-mail: gao.jianjun@shufe.edu.cn).}
	\thanks{Chengneng Jin is with the School of Information Managment and Engineering, Shanghai University of Finance and Economics, Shanghai 200433 China (e-mail: jin.chengneng@gmail.com).}
	\thanks{Yun Shi with School of Statistics, East China Normal University, Shanghai 200062, China (email: yshi@fem.ecnu.edu.cn)}
	\thanks{Xiangyu Cui, \textit{Corresonding auithor}, is with the School of Statistics and Management, Dishui Lake Advanced Finance Institute, Shanghai University of Finance and Economics, Shanghai 200433 China (e-mail: cui.xiangyu@mail.shufe.edu.cn).}
}

\maketitle

\begin{abstract}
Motivated by practical applications, we explore the constrained multi-period mean-variance portfolio selection problem within a market characterized by a dynamic factor model. This model captures predictability in asset returns driven by state variables and incorporates cone-type portfolio constraints that are crucial in practice. The model is broad enough to encompass various dynamic factor frameworks, including practical considerations such as no-short-selling and cardinality constraints. We derive a semi-analytical optimal solution using dynamic programming, revealing it as a piecewise linear feedback policy to wealth, with all factors embedded within the allocation vectors. Additionally, we demonstrate that the portfolio policies are determined by two specific stochastic processes resulting from the stochastic optimizations, for which we provide detailed algorithms. These processes reflect the investor's assessment of future investment opportunities and play a crucial role in characterizing the time consistency and efficiency of the optimal policy through the variance-optimal signed supermartingale measure of the market. We present numerical examples that illustrate the model's application in various settings. Using real market data, we investigate how the factors influence portfolio policies and demonstrate that incorporating the factor structure may enhance out-of-sample performance.
\end{abstract}


\section{Introduction}\label{sec:introduction}

\IEEEPARstart{T}he classic static mean-variance (MV) portfolio selection model, proposed by \cite{Markowitz1952Portfolio}, laid the foundation for modern portfolio theory. However, extending this static MV portfolio to dynamic multi-period settings is challenging due to the variance term not satisfying Bellman's principle of optimality. To overcome this difficulty, \cite{li2000optimal} and \cite{zhou2000continuous} considered a model with independent returns and developed a globally optimal portfolio policy by embedding the problem into an optimal control problem. Subsequent research extended multi-period MV (MMV) portfolio selection models in various directions, see, e.g., \cite{basak2010dynamic,cui2014optimal,gao2015time,StadenDangForsyth:2020}.

While theoretical studies on MMV portfolio models show promise, practical implementation faces challenges. On one hand, although the factor model is a fundamental building block of portfolio optimization, there are few studies that integrate factor models into MMV portfolio optimization (see \cite{garleanu2013dynamic,HeWangChen-TAC2015-portfolio}). In financial economics, various factor-based models have been proposed to characterize return predictability (see  \cite{campbell2002strategic,Tsay:book_ts_2010
,garleanu2013dynamic}). Despite some research considering return correlations in MMV portfolio models, existing works say little about integrating rich factor-based models into MMV portfolio selection. Another crucial issue in practical portfolio management is portfolio constraints, such as the no-short-selling constraint and sparse (cardinality) constraints. To address these issues, this paper studies the MMV portfolio optimization problem in a factor-based market featuring cone constraints. We assume that the returns of risky assets are determined by a set of observable state variables governed by their own dynamics. This setting includes the Markov-regime switching model (see \cite{hu2005constrained,costa2008generalized}), the linear factor model (see \cite{garleanu2013dynamic}), the GARCH-type model, and the GARCH-jump mixture model (see \cite{li2020multivariate}) as special cases. We also consider the cone constraint on portfolio decisions, which subsumes several practical constraints as special cases.

The paper makes several key contributions. Firstly, we derive semi-analytical portfolio policies for the factor-based MMV portfolio decision problem with general cone constraints. Theoretically, this problem is a constrained stochastic optimal control problem with a highly coupled vector state space (i.e., financial factors), which typically precludes an analytical solution in a general setting. However, due to its special structure, we demonstrate that the optimal policy is a piecewise linear feedback policy of the current wealth level, with the corresponding allocation vectors derived from two stochastic processes, termed Future Investment Opportunity (FIO) processes. We uncover a unique feature of separability between wealth and FIO processes, allowing these stochastic processes to be generated using the embedding method and dynamic programming. In the absence of portfolio constraints, we show that the FIO processes reduce to a Riccati equation-type recursion. Secondly, we provide numerical algorithms to compute the allocation vectors and FIO processes. These algorithms can be implemented by sequentially solving a series of cone-constrained stochastic optimization problems offline. Since these numerical methods require fitting the FIO process to discrete data points, we propose using neural networks for approximation, which achieves robust performance. Thirdly, we analyze the issue of time consistency in efficiency (TCIE) for the resulting optimal policy. The TCIE property, as proposed by \cite{cui2012better} and \cite{cui2017mean}, requires that the truncated pre-committed optimal policy remains mean-variance efficient for the time-truncated problem. With the help of the FIO processes, we derive the conditions that guarantee the TCIE property by using the Variance-Optimal Signed Supermartingale Measure (VSSM). This result extends the work of \cite{cui2017mean}, which assumes independence of returns and ignores the constraints. Finally, we provide numerical examples to illustrate the detailed solution procedure and conduct an empirical analysis using real market data based on the Fama-French five-factor model (see \cite{fama2015five}). This analysis investigates the impact of the factors on portfolio policy and demonstrates the superior performance of integrating the factor model into the MMV portfolio policy in terms of out-of-sample performance.

In the literature, several attempts have been made to extend the basic MMV model in \cite{li2000optimal} to more realistic market settings. Addressing various portfolio constraints, \cite{zhu2004risk} and  \cite{li2012multi} considered the bankruptcy prohibition constraint. \cite{gao2015time} explored the time-cardinality constraint, while \cite{li2002dynamic} and \cite{cui2014optimal} focused on the no-shorting constraint. \cite{hu2005constrained}, \cite{czichowsky2013cone}, \cite{cui2017mean}, and \cite{wu2018explicit} considered the cone-constrained model, and \cite{czichowsky2012convex}, and \cite{nystrup2019multi} explored general convex portfolio constraints. These studies either derived analytical portfolio policies or developed numerical solution approaches. However, due to complexity, this research typically ignores the factor structure in the market and assumes a simple assumption of the assets' returns. Similarly, the correlation of asset returns in MMV portfolio selection models has been explored. For example, \cite{yin2004markowitz}, and \cite{wu2012multi} investigated regime-switching markets, while \cite{costa2008generalized}, and \cite{yao2013uncertain} introduced jump terms. Additionally, \cite{GaoLiCuiWang:2015}, \cite{HeWangChen-TAC2015-portfolio}, and \cite{wu2018explicit} considered general serially correlated returns. Although these studies addressed certain structures of serially correlated returns, they rely on simplified assumptions about asset returns and often overlook portfolio constraints. These assumptions enable the derivation of the stochastic Riccati equation (see \cite{yin2004markowitz, GaoLiCuiWang:2015}), leading to explicit portfolio policies. However, when portfolio constraints are introduced, these methods cannot be applied directly. This paper aims to address this challenge.

To characterize assets' returns, there are plenty of financial econometric {models} in the finance literature. For example, the famous Fama-French three and five factor model (see \cite{fama1993common,fama2015five}) are often used to describe the excess returns of risky assets. The vector autoregression model (see \cite{campbell2002strategic}) captures the linear dynamic nature between returns and factors. Autoregressive conditional heteroskedasticity models (GARCH, \cite{bollerslev1986generalized}), and its various extensions, such as EGARCH (see \cite{nelson1991conditional}), GJR-GARCH (see \cite{breen1989economic}), are often used to describe the variance of excess returns. These rich time series models motivate us to consider how to unify these factor-based models MMV portfolio selection.

This paper contributes to the ongoing developments in MMV portfolio selection, with a primary focus on the pre-committed policy. Although this policy lacks time consistency, it has a notable advantage: when evaluated over the entire investment period, it is globally optimal compared to the time-consistent policy (see \cite{WangForsyth:2012IJTAF,Vigna2020}). Furthermore, \cite{StadenDangForsyth:2020} demonstrates that the pre-committed policy exhibits robust properties even when the model is misspecified. \cite{StadenDangForsyth:2020distribution} examines the distribution of terminal wealth generated by different MV policies, while \cite{cong2016pre} explores the relationship between the pre-committed and time-consistent policies. In addition to the pre-committed policy, time-consistent (or equilibrium) policies have also been extensively studied (see \cite{NicoleAnna,NiLiJiMiroslav-2020}).

The rest of paper is organized as follows. Section \ref{sec_model} presents the basic market setting and the constrained MMV portfolio selection models. Section \ref{sec_mv_policy} develops the optimal policies and provides numerical {algorithms} for different settings. Section \ref{sectime-consistency} investigates time consistency in efficiency (TCIE) property of our policy. In Section \ref{sec_empirical_analysis}, we conduct the empirical analysis based on the real market data. Section \ref{sec_conclusion} concludes paper and {discusses} the future research directions.

Notations: We use the boldface letter to denote vector or matrix. The notation $\1_{\cA}$ represents the indicator function, i.e., $\1_{\cA}=1$ if event $\cA$ is true and $\1_{\cA}=0$ otherwise. If matrix $\bm{A}$ is {positive} definite (positive semidefinite), we denote it as $\bm{A}\succ 0$ ($\bm{A}\succeq 0$). We use $\mathcal{N}(\bm{\mu}, \bm{\Sigma})$ to denote the multivariate normal distribution with $\bm{\mu}$ and $\bm{\Sigma}$ being the expected value and {variance}. For a problem $\cP$, we use $v(\cP)$ to denote its optimal objective value.  

\section{Market Model and Problem Formulation}\label{sec_model}

We consider a market comprising one risk-free asset and $N$ risky assets, with their prices {evolving over $T+1$ discrete time instances, labeled $t=0,\ldots,T-1(T?)$}. We use $r^0_{t+1}$ to denote the deterministic return of risk-free assets and use $\r_{t+1} \in \mathbb{R}^N$ to denote the excess random vector of $N$ risky assets in the $t$-th period $t$ for $t=0,\ldots, T-1$.\footnote{If $S_t^i$ is the price of the $i$-th risky asset at time $t$, the excess return is $r_{t+1}^i\triangleq (S_{t+1}^i - S_{t}^i)/S_{t}^i - r_{t+1}^0$ for $i=1,\ldots, N$ and the excess return vector is $\r_{t+1}\triangleq \big(r_{t+1}^1,\ldots, r_{t+1}^{N}\big)^{\top}$ for $t=0,\ldots,T-1$.} All the random variables/processes in this paper are defined on a complete probability space $\{\Omega,\{\mathcal{F}_t\}|_{t=0}^T, \mP\}$, where $\mP$ is the probability measure, $\{\cF_t\}|_{t=0}^T$ is the information set satisfying $\mathcal{F}_0=\{\emptyset, \Omega\}$, $\cF_s \subseteq \cF_t$ for $s\leq t$.\footnote{Our market model follows a similar setting as introduced by \cite{Follmerbook2016}, where a rigorous discrete-time market model is defined. The information set $\{\cF_t\}|_{t=0}^T$ is also referred to as the \textit{filtration}, with $\cF_t$ representing the $\sigma$-algebra that models the class of all events observable up to time $t$.}

Motivated by empirical studies highlighting the predictability of returns characterized by dynamic factor models (see, e.g., \cite{campbell2002strategic,AndrewTimmermann:2012,garleanu2013dynamic}), we assume the excess returns $\r_{t}$ satisfies the following factor model with $N_s$-dimensional state variables $\s_{t} \in \mathbb{R}^{N_s}$,
\begin{align}
\r_{t} = \f(t, \s_{t},\bm \epsilon_{t}),~~t=1,\ldots, T,\label{eqr}
\end{align}
where $\f(\cdot,\cdot,\cdot):\mathbb{R}\times \mathbb{R}^{N_s}\times \mathbb{R}^N \rightarrow \mathbb{R}^N$ is a deterministic vector function, $\bm \epsilon_{t} \in \mathbb{R}^N$ is an $\mathcal{F}_{t}$-measurable random vector. We also model the dynamics of the state variables as follows,
\begin{align}
 \s_{t} = \g(t,\s_{t-1},\bm \xi_{t}),~~t = 1,\ldots, T,\label{eqs}
\end{align}
where $\g(\cdot,\cdot,\cdot):\mR\times \mR^{N_s} \times \mR^{N_{\epsilon}}\rightarrow \mR^{N_s}$ is a deterministic vector function, $\bm \xi_{t}\in \mR^{N_{\xi}}$ is also an $\mathcal{F}_{t}$-measurable random vector. The two random vectors $\bm \epsilon_t$ and $\bm \xi_t$ may be correlated to each other. At the time $t$, the realizations of excess returns $\r_{t}$ and state variable $\s_t$ are observable, i.e., $\r_t \in \mathcal{F}_{t}$ and $\s_t \in \mathcal{F}_{t}$. This framework reveals that excess returns are influenced by both the uncertainty in the dynamics of state variables and the uncertainty between excess returns and state variables. 

In the subsequent section, the notations $\E[\cdot|\cF_t]$ ($\E[\cdot]$)  and $\Var[\cdot|\cF_t]$ ($\Var[\cdot]$) denote the conditional (unconditional) expectation and variance with respect to $\cF_t$ for $t=0,\ldots,T-1$.  To emphasize the dependency on the state variable $\s_t$, we also use $\E[\cdot|\s_t]$ instead of $\E[\cdot|\cF_t]$ when there is no ambiguity.

The dynamics of excess returns in (\ref{eqr}) and (\ref{eqs}) are sufficiently broad to encompass various existing models addressing return predictability in financial literature as specific instances. 

Model (\ref{eqr}) and (\ref{eqs}) naturally encompass the Markov chain regime-switching model in  \cite{yin2004markowitz, ccakmak2006portfolio}, and  \cite{costa2008generalized} as a special case. Specifically, we may assume $s_t \in\{1, 2,\ldots, M\}$ follow a discrete-time Markov chain with $M$ states for $t=0,\ldots, T$. The state equation (\ref{eqs}) can be characterized by the one-step transition probability, i.e., the state transits from $s_t=i$ to $s_{t+1}=j$ with probability $P_{i,j} \triangleq \mP(s_{t+1}=j | s_{t}=i)$ for all $i,j=1,\ldots, M$.\footnote{This setting aligns with (\ref{eqs}) by treating $\xi_{t}$ as a dummy random variable representing the state transition. \cite{Bertsakas:book:optimalcontrol} (vol.1) details reformulating the Markov chain-based dynamics to the system-based state equation.} We then let $\bm{c}(s_t)$ and $\bm{\Sigma}(s_t)$ be the parameters governed by $s_t$, enabling us to represent (\ref{eqr}) as
\begin{align}
\r_{t} = \bm c(s_{t}) + \bm{\Sigma}^{\frac{1}{2}}(s_{t}) \bm{\epsilon}_{t},~~t=0,1,\ldots, T-1, \label{def_rt_markov}
\end{align}
where $\bm{\epsilon}_t \sim \mathcal{N}( \0, \bm{I})$. In this case, $\r_t$ follows a multivariate normal distribution whose mean vector, $c(s_t)$, and covariance matrix $\bm{\Sigma}(s_{t})$ are governed by the state variable $s_t$.

We may regard (\ref{eqs}) as the linear dynamic factor model of the assets' returns.\footnote{One typical example is the three-factor model (see \cite{fama1993common}). Specifically, we choose the excess returns of the market portfolio (Mkt), the SMB (Small Minus Big) portfolio, and the HML (High Minus Low) portfolio as state variables, i.e., $\s_t = \big(\textrm{Mkt}_t, \textrm{SMB}_t, \textrm{HML}_t\big)^{\top}$.} Combining the mean-reverting dynamics of the factors (see \cite{garleanu2013dynamic}), (\ref{eqr}) and (\ref{eqs}) become
\begin{align*}
\begin{dcases}
\r_{t} = \bm \alpha + \bm B \s_{t}  +\bm \epsilon_{t},\\
\s_{t} = (\bm I-\bm \Phi) \s_{t-1} +\bm \xi_{t},
\end{dcases}
\end{align*}
for $t=1,\ldots, T,$ where $\bm{\alpha}$, $\bm{B}$ and $\bm{\Phi}$ are parameters in proper dimension; and $\bm\epsilon_{t}\sim \mathcal{N}(\0,\bm \Sigma_{\bm\epsilon})$ and $\bm{\xi}_{t}\sim \mathcal{N}\big( \0, \bm{\Sigma}_{\bm{\xi}} \big)$ are independent random vectors over time. 

State dynamics (\ref{eqr}) and (\ref{eqs}) can be also  used to represent GARCH-type volatility models. Let $\bm{\mu}$ be the mean of the excess return. We may regard the conditional covariance matrix $\bm{H}_t$ and the error vector $\bm{\eta}_t$ as the state variables, then it has
\begin{align*}
\begin{dcases}
\r_{t} = \bm \mu + \bm H_{t}^{\frac{1}{2}} \bm \eta_{t},\\
\bm{H}_{t} = \bm{C}^{\top} \bm{C} + \bm{A}^{\top} \bm{\eta}_{t-1}\bm{\eta}_{t-1}^{\top}\bm{A} + \bm{B}^{\top} \bm H_{t-1} \bm B,\\
\bm \eta_{t} = \bm \xi_t,\\
\end{dcases}
\end{align*}
for $t=1,\ldots, T,$ where $\bm{C}$, $\bm{A}$, and $\bm{B}$ are the {parameters}; and $\bm H_{t}^{\frac{1}{2}}$ is the Cholesky decomposition of the conditional covariance matrix $\bm H_{t}$ and $\bm{\xi}_{t}\sim \mathcal{N}(\0,\bm{I})$. Such a model represents one important multivariate GARCH model known as the BEKK-GARCH model  (see \cite{engle1995multivariate}). Besides this model, if we regard $\bm{H}_t$ as the factor and regard $\big(\bm\eta_t, \bm Y_t, \bm B_t\big)^{\top}$ as an aggregated random vector representing by $\bm\epsilon_t$ in (\ref{eqr}), the dynamics (\ref{eqr}) and (\ref{eqs}) can represent a multivariate GARCH-Jump mixture model (see \cite{li2020multivariate}).

It is worth mentioning that the state dynamics (\ref{eqr}) and (\ref{eqs}) also cover the market with independent and identically distributed random returns (see \cite{li2000optimal,cui2014optimal,Pun:Automatic_2022_transaction}). The dynamic factor model (\ref{eqr}) and (\ref{eqs}) also have some limitations, e.g., they can not include the stochastic volatility model (see \cite{harvey1994multivariate}) as its special case as there is a latent variable that can not be observed at time $t$.

Given the market model, we consider the multi-period MV portfolio selection problem. The investor enters the market with the initial wealth $x_0$ and allocates the wealth in $N$ risky assets each time period. We use $\bm{\pi}_t\in \mR^N$ to denote the portfolio decision representing the wealth allocation at time $t$, $t=0,\ldots, T-1$. Investor's wealth at time $t$ is denoted by $x_t$, which follows the following dynamic (see \cite{li2000optimal}),
\begin{align}
x_{t+1} = r^0_{t+1} x_t + \r_{t+1}^{\top}\bm{\pi}_t, ~ t=0,1,\dots,T-1.\label{def_xt}
\end{align} 
We further require that the admissible portfolio policy $\bm{\pi}_t$ in period $t$ should be $\mathcal{F}_t$-measurable, i.e., {$\bm{\pi}_t$} should only depend on the information up to time $t$.

In practical portfolio management, the portfolio is usually restricted by various constraints. In this work, we assume the portfolio policy $\bm{\pi}_t$ are confined in a deterministic cone denoted by $\cA_t$ for $t=0,\ldots,T-1$. Such a cone constraint may not necessarily be convex. We focus on three types of constraints: (I) \textbf{The No-shorting Constraint}: We set $\cA_t=\{\bm{\pi}\in \mR^{N}~|~\bm \pi \geq \0 \}$, which represents the no-shorting constraint. Besides some markets that physically prohibits short-selling, the no-shorting constraint is usually added in the portfolio optimization models to enhance the out-of-sample performance (see \cite{Jagannathan:JOF2003}). Clearly, when $\cA_t =\mR^{N}$, it means the portfolio is not constrained. (II) \textbf{Cardinality Constraint}: Such a constraint (also known as the sparse constraint) controls the number of active risky assets in a large assets pool (see \cite{GaoLi:card2013,Bertsimas:2022}). Let $\textrm{sign}(\cdot):\mR \rightarrow \mR$ be sign function, i.e., $\textrm{sign}(x)=1$ if $x>0$, $\textrm{sign}(x)=-1$ if $x<0$, and $\textrm{sign}(x)=0$ if $x=0$. Then, the cardinality constraint is 
$\mathcal{A}_t =\big\{\bm{\pi}=(\pi_1,\ldots, \pi_N)^{\top}~\big|~\sum_{i=1}^N |\textrm{sign}(\pi_i)|\leq N_{\textrm{active}} \big\}$,
where $N_{\textrm{active}} \leq N$ is the number of active assets. If $\bm{\pi}_t \in \cA_t$, it means that the investor {selects no more than} $N_{\textrm{active}}$ assets from total $N$ assets to trade in period $t$. Clearly, for any $\pi_t \in \cA_t$ it has $\alpha \cdot \pi_t \in \cA_t$ for any $\alpha>0$, which implies $\cA_t$ is a cone. (III) \textbf{Linear Cone Constraint} We may set $\cA_t=\{\bm{\pi}\in\mR^N~|~ \bm{A} \bm{\pi}\geq \0,~~\bm{A}\in \mR^{K\times N}\}$ which represents the noshorting for the combination of portfolio. For example, considering the simple long-short portfolio, $\bm{\pi}=\{\pi_1,\ldots,\pi_N\}$, if $\bm{A}=\bm{e}^{\top}$, the constraint $\bm{e}^{\top}\bm{\pi} = \sum_{i=1}^N \pi_{i} \geq 0 $ means that the investor's total risky position should be nonnegative. Note that, as the intersection of the cones is also a cone, if there are different types of cone constraints, it still fits our setting. 

Before we introduce the portfolio selection problem, we impose the following assumption for the stochastic market. 
\begin{assumption}\label{ass1}
At the beginning of period $t-1$, with knowing the state variable $\s_{t-1}$, the excess returns $\r_t$ of $N$ risky assets are linearly independent to each other for all $t=1,\ldots,T$. 
\end{assumption}
The Assumption \ref{ass1} is similar to the no arbitrage requirement in the finance literature. Let $\Cov[\r_t|\s_{t-1}]$ be the {conditional} covariance matrix of $\r_t$ given state $\s_{t-1}$. Under the assumption, it has $\Cov[\r_t|\s_{t-1}]\succ 0$, which implies that there does not exist the portfolio policy such that $\Var[\r_{t}^{\top} \bm{\pi}_{t-1}|\s_{t-1}]$$=$$\bm{\pi}_{t-1}^{\top}\Cov[ \r_t | \s_{t-1} ] \bm{\pi}_{t-1} = 0$ and $\E[\r_{t}^{\top} \bm{\pi}_{t-1} |\s_{t-1}] >0$. In a plain language, there does {not} exist a portfolio policy, which can achieve a deterministic positive expected portfolio return with zero variance (risk) in that period. 

Under the proposed market setting and portfolio constraint, the investor considers the following cone {constrained} MMV portfolio {optimization} problem,
\begin{align*}
\cP_{\mv}^1(\lambda)\min_{\{\bm{\pi}_t \in \mathcal{A}_t \}|_{t=0}^{T-1}}~& \Var[x_T] - 2\lambda  \E[x_T] \notag\\
\mbox{s.t.}~~~&\textrm{$\{x_t, \bm{\pi}_t,\r_t,\s_t \}$ {satisfies} (\ref{eqr}), (\ref{eqs})},(\ref{def_xt}),
\end{align*}
where $\lambda \geq 0$ is the risk aversion parameter. Besides the above MV-weighted summation formulation, the problem can be also formulated as variance minimization problem with given target expected portfolio return
\begin{align*}
\cP_{\mv}^2(x_{\tg})\min_{\{\bm{\pi}_t \in \mathcal{A}_t \}|_{t=0}^{T-1}}\quad &\Var[x_T] \notag\\
\mbox{s.t.}~~~&\E[x_T] = x_{\tg},\\
           ~&\textrm{$\{x_t, \bm{\pi}_t,\r_t,\s_t \}$ {satisfies} (\ref{eqr}), (\ref{eqs})},(\ref{def_xt}),
\end{align*}
where $x_{\tg}$ is the given target terminal wealth level. We denote $\rho_t \triangleq \prod_{k = t}^{T-1} r^0_{k+1}$ for $t=0,\ldots,T-1$ and $\rho_T \triangleq 1$ which represents the cumulative {risk-free} return starting from time $t$. We assume $x_{\tg}\geq \rho_0 x_0$, i.e., target expected terminal wealth should be at least larger than the pure investment in risk free asset.

\section{Optimal Portfolio Policies and Numerical Methods}\label{sec_mv_policy}

\subsection{Optimal Portfolio Policies}\label{ssepolicy}
The variance term in the objective functions of problems $\cP^1_{\mv}(\lambda)$ and $\cP^2_{\mv}(x_{\tg})$ does not satisfy the separation property in the sense of Bellman's principle of optimality, preventing these problems from being solved directly through dynamic programming. To address this, we employ the embedding method proposed by \cite{li2000optimal}. Specifically, we first solve a parameterized auxiliary problem $\mathcal{L}(\lambda,b)$, which is a linear-quadratic stochastic optimal control problem. We then use the solution of this auxiliary problem to retrieve the solution to the original problem. The auxiliary problem  $\mathcal{L}(\lambda,b)$ is define as follows:
\begin{align}
\mathcal{L}(\lambda,b)  \min_{\{ \bm{\pi}_t \in \mathcal{A}_t  \}|_{t=0}^{T-1}}\quad &\E\big[ (x_T-b)^2 \big]  - 2 \lambda  \E[x_T],\label{L_obj}\\
\mbox{s.t.}~~~&\textrm{$\{x_t, \bm{\pi}_t,\r_t,\s_t \}$ {satisfies} (\ref{eqr}), (\ref{eqs})},(\ref{def_xt}),\notag
\end{align}
where $b\in \mR$ is a {parameter}. Clearly, when the parameter $b$ happens to be $b=\E[x_T]$, problem $\mathcal{L}(\lambda,b)$ becomes the original problem $\cP^1_{\mv}(\lambda)$. In this case, the solution of the auxiliary problem $\mathcal{L}(\lambda,b)$ also solves problem $\cP_{\mv}^1(\lambda)$. Different from $\cP_{\mv}^1(\lambda)$, the auxiliary problem $\mathcal{L}(\lambda, b)$ can be solved by dynamic programming. 

We may derive the solution by using a more concise formulation, i.e., we replace the wealth $x_t$ by a shifted wealth as follows,
\begin{align}
 y_t \triangleq  x_t - \frac{b + \lambda}{\rho_t},~~t=0,\ldots, T. \label{def_yt_xt}
\end{align}
Note that, as $\rho_T =1$, it has $y_T = x_T - (b + \lambda)$. After replacing the wealth by shifted wealth $y_t$, we may add term $2\lambda  b + \lambda^2 $ into the objective function (\ref{L_obj}), which does affect optimal solution. Such a modification enables us to simplify the problem $\mathcal{L}(\lambda,b)$ as follows,
\begin{align}
\cA(\lambda,b) \quad \min_{\{ \bm{\pi}_t \in \mathcal{A}_t \}|_{t=0}^{T-1}}\quad &\E\big[ y_T^2\big] =\E[(x_T-(b+\lambda))^2 ],\notag\\
\mbox{s.t.}\quad &~y_{t+1} = r^0_{t+1} y_t +\r_{t+1}^{\top}\bm{\pi}_t,\label{eqyt}\\
&~\textrm{$\{\r_t,\s_t \}$ {satisfies} (\ref{eqr}), (\ref{eqs})}.\notag
\end{align}

To solve the auxiliary problem $\cA(\lambda,b)$, we introduce two one-dimensional $\mathcal{F}_t$-measurable stochastic processes, $\{d_t^+(\s_t)\}|_{t=0}^{T}$, $\{d_t^-(\s_t)\}|_{t=0}^{T}$, which are defined recursively for $t=T-1,T-2,\dots,1$,
\begin{align}
d_t^-(\s_t) &= \min_{\k_t\in\cA_t} \E\big[ \big(1 -\r^{\top}_{t+1} \bm{k}_t\big)^2 \big( d_{t+1}^-(\s_{t+1})
 \1_{\{\r^{\top}_{t+1}\bm{k}_t \leq1 \}}\notag\\
            &~~~~~~~~~~+ d_{t+1}^+(\s_{t+1}) \1_{\{\r^{\top}_{t+1}\bm{k}_t > 1 \}}\big)\big|\s_t\big],\label{eqd_t_minus}\\
d_t^+(\s_t) &=\min_{\k_t\in\cA_t} \E\big[ \big(1+\r^{\top}_{t+1}\bm{k}_t\big)^2 \big( d_{t+1}^-(\s_{t+1}) 
\1_{\{\bm{r^{\top}}_{t+1}\bm{k}_t \geq -1 \}}\notag \\
            &~~~~~~~~~~+d_{t+1}^+(\s_{t+1}) \1_{\{\r^{\top}_{t+1}\bm{k}_t <-1 \}}\big)\big|\s_t\big],\label{eqd_t_plus}
\end{align}
where the boundary {condition} are $d_{T}^-(\s_T)= d_T^+(\s_T) =1$.\footnote{As $\s_t$ is random variable, the ``minimization '' defined in problems (\ref{eqd_t_minus}) and (\ref{eqd_t_plus}) should be ``essential infimum'', i.e., we want to minimize those function values with the nonzero measure. However, due to the readability, we {still} use the `minimization' in these problems.} Associated with the two {problems (\ref{eqd_t_minus})} and (\ref{eqd_t_plus}), we denote the minimizers of {problems} (\ref{eqd_t_minus}) and (\ref{eqd_t_plus}) by $\{\k_t^+(\s_t)\}|_{t=0}^{T-1}$ and $\{\k_t^-(\s_t)\}|_{t=0}^{T-1}$, respectively.

Note that $\{\k_t^+(\s_t)\}|_{t=0}^{T-1}$, $\{\k_t^-(\s_t)\}|_{t=0}^{T-1}$ are two $N$-dimensional $\mathcal{F}_t$-measurable stochastic processes. In the above formulations, the distribution of $\r_{t+1}$ conditional on $\s_t$ is fully characterized by the dynamic equations in (\ref{eqr}) and (\ref{eqs}). The processes $\{d_t^-(\s_t)\}_{t=0}^{T}$ and $\{d_t^+(\s_t)\}_{t=0}^T$ have the following properties.
\begin{lemma}\label{lem-inequ}
At any time $t=0,1,\ldots, T-1$, the random variables $d_t^-(\s_t)$ and $d_t^+(\s_t)$ defined in (\ref{eqd_t_minus}) and (\ref{eqd_t_plus}) satisfy the following {inequalities},  $0< d_t^-(\s_{t}) \leq 1$, $0< d_t^+(\s_{t}) \leq 1$, $d_t^-(\s_{t})\leq \E[d_{t+1}^-(\s_{t+1}) |\s_t]$, and $d_t^+\leq \E[d_{t+1}^+(\s_{t+1})|\s_t]$ almost surely.
\end{lemma}

\proof Note that, under Assumption \ref{ass1}, for any probability measure $\mQ$ equivalent to probability measure $\mP$, the excess returns of $N$ risky assets are also linearly independent. Thus, under probability measure $\mQ$, we have $\mbox{Cov}^{\mQ}(\r_t|\s_{t-1})\succ 0$. According to Schur Complement Lemma, we have 
\begin{align*}
\begin{pmatrix}
1 & \E^{\mQ}[\r_t^{\top}|\s_{t-1}]\\
\E^{\mQ}[\r_t|\s_{t-1}] & \E^{\mQ}[\r_t\r_t^{\top}|\s_{t-1}]
\end{pmatrix}\succ 0,
\end{align*}
which further implies that $\E^{\mQ}[\r_t\r_t^{\top}|\s_{t-1}] \succ 0$ and $1- \E^{\mQ}[\r_t^{\top}|\s_{t-1}](\E^{\mQ}[\r_t\r_t^{\top} |\s_{t-1}])^{-1}\E^{\mQ}[\r_t |\s_{t-1}]>0$. We then prove the proposition. Since it has $d_T^-(\s_T) =d_T^+(\s_T)=1$, we only need to prove that when $0<d_{t+1}^-(\s_{t+1})\leq 1$ and $0<d_{t+1}^+(\s_{t+1})\leq 1$ hold, the statements are true for time $t$. We focus on the optimization problem (\ref{eqd_t_minus}) and $d_t^-(\s_t)$. Denote the objective function of optimization problem in (\ref{eqd_t_minus}) as 
\begin{align*}
& h_t^-(\k_t;\s_t) = \E \big[ \big(1 -\r^{\top}_{t+1} \bm{k}_t\big)^2 \big( d_{t+1}^-(\s_{t+1}) \1_{\{\r^{\top}_{t+1}\bm{k}_t \leq1 \}}\\
&+  d_{t+1}^+(\s_{t+1})  \1_{\{\r^{\top}_{t+1}\bm{k}_t > 1 \}}\big)|\s_t\big].
\end{align*}
We can compute the first-order and second-order derivatives of the objective function as follows,
\begin{align*}
&\nabla_{\k_t} h_t^-(\k_t;\s_t) =2 \E\big[ \big(\r_{t+1}  -\r_{t+1}\r^{\top}_{t+1}\bm{k}_t\big) \big( d_{t+1}^-(\s_{t+1})\\
&~~\times \1_{\{\r^{\top}_{t+1}\bm{k}_t \leq1 \}}+  d_{t+1}^+(\s_{t+1})\1_{\{\r^{\top}_{t+1}\bm{k}_t > 1 \}}\big)\big|\s_t\big],\\
&\nabla_{\k_t}^2  h_{t}^-(\k_{t};\s_{t}) = 2\E\big[ \r_{t+1}\r^{\top}_{t+1} \big( d_{t+1}^-(\s_{t+1}) \1_{\{\r^{\top}_{t+1}\bm{k}_t \leq1 \}}\\
&~+d_{t+1}^+(\s_{t+1})  \1_{\{\r^{\top}_{t+1}\bm{k}_t > 1 \}}\big) \big| \s_t\big]\\
&~\succeq 2\E\big[ \min\{ d_{t+1}^-(\s_{t+1}), d_{t+1}^+(\s_{t+1}) \} \r_{t+1}\r^{\top}_{t+1} \big| \s_t\big]\\
&~=2 \E\big[ \min\{ d_{t+1}^-(\s_{t+1}), d_{t+1}^+(\s_{t+1}) \} \big| \s_t\big]\E^{\mQ}\big[  \r_{t+1}\r^{\top}_{t+1}|\s_t\big]\\
&~ \succ 0,
\end{align*}
where the probability measure $\mQ$ is defined through the density of $\mQ$ with respect to $\mP$ as 
\begin{align*}
\frac{\mathrm{d} \mQ}{\mathrm{d} \mP} = \frac{ \min\{ d_{t+1}^-(\s_{t+1}), d_{t+1}^+(\s_{t+1}) \} }{\E[ \min \{d_{t+1}^-(\s_{t+1}), d_{t+1}^+(\s_{t+1}) \}|\s_t]}. 
\end{align*}
In the following part, we first prove result when the cone $\cA_{t}$ is convex (i.e., the cases (I) and (III) in Section \ref{sec_model}) then we extend the result to the cardinality constraint (i.e., case (II) in Section \ref{sec_model}). We then focus on $d_t^-(\s_t)$. Form the definition of $d_t^-(\s_t)$, it has
\begin{align*}
&d_t^-(\s_t) =\min_{\k_t\in\cA_t} h_t^-(\k_t;\s_t) \geq \min_{\k_t\in\cA_t} \E\big[ \min\{d_{t+1}^-(\s_{t+1}),\\
&d_{t+1}^+(\s_{t+1}) \}\big(1 -\r^{\top}_{t+1}\bm{k}_t\big)^2 \big|\s_t\big] \geq \min_{\k_t\in\mathbb{R}_N}\E\big[ \min\{d_{t+1}^-(\s_{t+1}), \\
&d_{t+1}^+(\s_{t+1}) \}|\s_t\big]\E^{\mQ}\big[\big(1 -\r^{\top}_{t+1}\k_t\big)^2|\s_t\big]\\
&= \E\big[ \min\{d_{t+1}^-(\s_{t+1}), d_{t+1}^+(\s_{t+1}) \}|\s_t\big]\big[1- \E^{\mQ}[\r_{t+1}^{\top}|\s_{t}] \\
&~~~~~~~~~\times(\E^{\mQ}[\r_{t+1}\r_{t+1}^{\top} \big|\s_{t}])^{-1}  \E^{\mQ}[\r_{t+1} \big |\s_{t}]\big]>0. 
\end{align*}
Furthermore, according to Theorem 27.4 in \cite{rockafellar1970convex}, $\k_t^{-}$ is optimal if and only if
$\big(\nabla_{\k_t}
h_t^{-}(\k_t^{-};\s_t)\big)^{\top}(\k_t-\k_t^{-})\geq 0$, $\forall~ \k_t\in \cA_t$
, which implies $\big(\nabla_{\k_t} h_t^{-}(\k_t^{-};\s_t)\big)^{\top} (\alpha\k_t^{-}-\k_t^{-})\geq 0$ for $\alpha>0$ and
$\big(\nabla_{\k_t}h_t^{-}(\k_t^{-};\s_t)\big)^{\top}\k_t^{-}=0$. Then, it has 
\begin{align*}
&d_t^-(\s_t)  = h_t^-(\k_t^-;\s_t)\\
&= \E\big[ \big(1 -(\k_t^-)^{\top} \r_{t+1}\r^{\top}_{t+1}\bm{k}_t^-\big) \big( d_{t+1}^-(\s_{t+1})\1_{\{\r^{\top}_{t+1}\bm{k}_t^- \leq1 \}}\\
&+  d_{t+1}^+(\s_{t+1})\1_{\{\r^{\top}_{t+1}\bm{k}_t^- > 1 \}}\big)|\s_t\big]
- \big(\nabla_{\k_t}
h_t^{-}(\k_t^{-};\s_t)\big)^{\top}\k_t^{-}\\
& \leq  \E\big[ \big(1 -(\k_t^-)^{\top} \r_{t+1}\r^{\top}_{t+1}\bm{k}_t^-\big)  d_{t+1}^-(\s_{t+1}) 1_{\{\r^{\top}_{t+1}\bm{k}_t^- \leq1 \}}|\s_t\big] \\
&\leq  \E\big[ d_{t+1}^-(\s_{t+1}) \1_{\{\r^{\top}_{t+1}\bm{k}_t^- \leq1 \}}|\s_t\big]
\leq \E\big[ d_{t+1}^-(\s_{t+1}) |\s_t\big]\leq 1.
\end{align*}
The result for $d_t^+(\s_t)$ can be proved in the similar fashion. As for the case when $\cA_t$ is the cardinality constraint, we may apply the above argument for each feasible sparse solution. Note that, given a feasible sparsity of the portfolio, some element of portfolio $\bm \pi_t$ are fixed to zero. Thus, we may remove all the coefficients related to those zero element, then the resulted problem is similar to the case with constraints are convex cone.  
\endproof

Lemma \ref{lem-inequ} shows that the processes $\{d_t^-(\s_t)\}|_{t=0}^{T}$ and $\{d_t^+(\s_t)\}|_{t=0}^T$ are bounded submartigngale, i.e., the average values of $d_t^-(\s_t)$ and $d_t^+(\s_t)$ increase as time goes by. Using this lemma, we derive the optimal policy of $\cA(\lambda,b)$ in the following proposition. 
\begin{proposition}\label{prop-auxilary}
The optimal portfolio policy of the auxiliary problem $\cA(\lambda,b)$ at period $t$ is given by
\begin{align}
\bm{\pi}_{t}^*(y_t,\s_t,b) &= r_{t+1}^0 ( -\bm{k}_t^-(\s_t) y_t \1_{\{y_t\leq 0\} }
                    +\bm{k}_t^+(\s_t)  y_t  \1_{ \{y_t>0\}}) \label{eqpolicy_problem_A}                     
\end{align}
for $t=0,\ldots,T-1$. Moreover, given $y_t$ at time $t$, the optimal value function defined by the following problem
\begin{align}
\begin{dcases}
J_t(y_t, \s_t ,b ) \triangleq  \min_{ \{ \bm{\pi}_k\in \cA_k \}_{k=t}^{T-1} }~~ \E[ y_T^2|\s_t]\\
\quad \quad \quad \quad \quad \mbox{s.t.}~~\textrm{$\{\r_k,\s_k, y_k \}$ satisfies (\ref{eqr}), (\ref{eqs}), (\ref{eqyt})}, 
\end{dcases}\label{def_prob_A_t}
\end{align}
for $t=0,\ldots, T-1$, is given by 
\begin{align}
J_t(y_{t},\s_t, b) &= \rho^2_{t} d_{t}^-(\s_t)  y_{t}^2 \1_{\{y_t\leq 0\}}
+ \rho^2_{t}  d_{t}^+(\s_t) y_{t}^2  \1_{\{y_t> 0\}}.\label{def_J_t}
\end{align} 
\end{proposition}
\proof At time $t$, the value function of auxiliary problem $\cA(\lambda,b)$ is defined by (\ref{def_prob_A_t}). To simplify the notation, we use $\E_t[\cdot]=\E[\cdot|\mathcal{F}_t]$. The Bellman equation of value function is
\begin{align*}
J_t(y_t,\s_t,b) = \min_{\bm{\pi}_t\in \cA_t} ~\E_{t} [J_{t+1}(y_{t+1},\s_{t+1},b)].
\end{align*}
Using the induction method, we show that the value function at time $t$ is in the following form,
\begin{align}
&J_t(y_t,\s_t, b) = \rho_{t}^{2} ( d_{t}^-(\s_t)   y_{t}^2\1_{\{y_{t} \leq 0\}} 
+ d_t^+(\s_t) y_{t}^2 \1_{\{y_{t}> 0\}} {)}.\label{induction_claim}
\end{align}
At time $T$, the claim (\ref{induction_claim}) is true, as it is $d_T^-(\s_T) =d_T^+(\s_T) = 1$ and $J_T(y_T,\s_T, b) = y_T^2$.  We then assume that the claim is true at $t+1$, i.e., it is $J_{t+1}(y_{t+1},\s_{t+1}, b)$ $=$ $d_{t+1}^-(\s_{t+1})\rho_{t+1}^{2} y_{t+1}^2 \1_{\{y_{t+1} \leq 0\}}$$+$$d_{t+1}^+(\s_{t+1}) \rho_{t+1}^{2} y_{t+1}^2 \1_{\{y_{t+1}> 0\}}$.
Then, we prove that the claim (\ref{induction_claim}) also holds at time $t$. Note that it has
\begin{align*}
&\E_t[J_{t+1}(y_{t+1},\s_{t+1}, b)]\\
=&  \E_t\Big[ \rho_{t+1}^{2} (r_{t+1}^0 y_{t}+\r_{t+1}^{\top} \bm \pi_t)^2 \big( d_{t+1}^-(\s_{t+1})  \1_{\{r_{t+1}^0y_{t} + \r_{t+1}^{\top}\bm \pi_t \leq 0\}}\\
&+ d_{t+1}^+(\s_{t+1}) \1_{\{r_{t+1}^0y_{t} + \r_{t+1}^{\top}\bm \pi_t> 0\}}\big) \Big].
\end{align*}
When $y_t<0$, we set $\bm \pi_t = - \bm k_t r_{t+1}^0 y_t$ where $\k_t$ is the newly defined decision variable. Then, the cone constraint $\bm \pi_t \in \cA_t$ becomes $\bm k_t \in\cA_t$. The optimization problem in Bellman equation becomes,
\begin{align*}
&\min_{\bm{\pi}_t\in \cA_t} ~\E_{t} [J_{t+1}(y_{t+1},\s_{t+1},b)]\\
=&\rho_t^2 y_t^2  \min_{\k_t\in \cA_t} ~\E\big[ \big(1 -\r^{\top}_{t+1}\bm{k}_t\big)^2  \big( d_{t+1}^- (\s_{t+1}) \1_{\{\r^{\top}_{t+1}\bm{k}_t \leq 1 \}}\\
&+  d_{t+1}^+ (\s_{t+1})  \1_{\{\r^{\top}_{t+1}\bm{k}_t > 1 \}}\big)|\s_t\big] .
\end{align*}
Solving the above problem gives the solution $\k^-_t(\s_t)$. Then the solution of value function is $\bm\pi_t^-(\s_t)$ $=$ $- \bm k_t^-(\s_t)r_{t+1}^0 y_t$. Substituting $\bm\pi_t^-(\s_t)$ into $ \E_t[J_{t+1}(y_{t+1},\s_t,b)]$, we obtain the value function at time $t$ as $J_t(y_t,\s_t,b) = d_t^-(\s_t) \rho_{t}^{2}  y_t^2$. When $y_t>0$, we apply the similar argument by setting $\bm \pi_t =  \bm k_t r_{t+1}^0 y_t$ for some $\k_t$. As the cone constraint $\bm \pi_t \in \cA_t$ implies $\bm k_t \in\cA_t$, the Bellman equation becomes,
\begin{align*}
&\min_{\bm{\pi}_t\in \cA_t} ~\E_{t} [J_{t+1}(y_{t+1},\s_{t+1})]\\
=&\rho_t^2 y_t^2 \min_{\k_t\in \cA_t}
\E\big[ \big(1+\r^{\top}_{t+1}\k_t\big)^2 \big( d_{t+1}^-(\s_{t+1}) 1_{\{\r^{\top}_{t+1}\bm{k}_t \geq -1 \}}\\
&+d_{t+1}^+(\s_{t+1}) 1_{\{\r^{\top}_{t+1}\bm{k}_t <-1 \}}\big)|\s_t\big] .
\end{align*}
Let $\k^+_t(\s_t)$ be the solution of the above problem. Then the optimal policy of the value function is $\bm{\pi}_t^+(\s_t) = \k_t^+(\s_t) r_{t+1}^0  y_t$. Substituting $\bm\pi_t^+(\s_t)$ into $ \E_t[J_{t+1}(y_{t+1},\s_{t+1})]$ yields the value function as $J_t(y_t,\s_t,b ) =  d_t^+(\s_t) \rho_{t}^{2}  y_t^2$.

When $y_t=0$,  the conditional expectation becomes 
\begin{align*}
&\E_t[J_{t+1}(y_{t+1},\s_{t+1},b)] \\
=&\E_t[ \rho_{t+1}^{2} ( \r_{t+1}^{\top} \bm \pi_t)^2 \big( d_{t+1}^-(\s_{t+1})\1_{\{\r_{t+1}^{\top} \bm \pi_t \leq 0\}}\\
&+ d_{t+1}^+(\s_{t+1})\1_{\{ \r_{t+1}^{\top}\bm \pi_t> 0\}})].
\end{align*}
We can see that $\bm \pi^* = \0$ is the optimal solution with $0$ being the objective value. Combining all the above cases, we obtain the value function at time $t$ as in claim (\ref{induction_claim}). \endproof

In Proposition \ref{prop-auxilary}, problem (\ref{def_prob_A_t}) is a truncated problem of the problem $\mathcal{L}(\lambda,b)$, which starts from time $t$. Thus, by setting $t=0$, the objective value of problem  $\mathcal{A}(\lambda,b)$ is 
\begin{align}
&v(\mathcal{A}(\lambda, b)) \notag \\
= &J_0(y_{0},\s_0, b) \notag \\
    =& \rho^2_{0}  d_{0}^-(\s_0)  y_{0}^2  \1_{\{y_0\leq 0\}}  
    + \rho^2_{0}  d_{0}^+(\s_0) y_{0}^2  \1_{\{y_0> 0\}}.\label{def_J_0}
\end{align} 
Recall that problems $\mathcal{A}(\lambda, b)$ and $\mathcal{L}(\lambda, b)$ merely alter the objective value by $2\lambda b + \lambda^2$. Consequently, the optimal policy for problem $\mathcal{L}(\lambda, b)$ is also determined by (\ref{eqpolicy_problem_A}), with the substitution of $y_t$ with $x_t$ in (\ref{def_yt_xt}).

By using the solution of the auxiliary problem $\mathcal{A}(\lambda,b)$, we may develop the solution for {original} problem $\cP^1_{\mv}(\lambda)$ as follows

\begin{theorem}\label{thm_policy_P1}
The optimal portfolio policy of problem $\cP_{\mv}^1(\lambda)$ is 
\begin{align}
&\bm{\pi}^{*}_{t}(x_t,\s_t)\notag \\
=&-\bm{k}_t^-(\s_t) \big( r_{t+1}^0 x_t - \frac{\rho_0x_0d_0^-(\s_0) + \lambda}{\rho_{t+1} d_0^-(\s_0)} \big)\1_{ \{\rho_t x_t \leq  \rho_0x_0 +\frac{\lambda}{d_0^-(\s_0)}\} }\notag \\
&+ \bm{k}_t^+(\s_t)\big(  r_{t+1}^0 x_t - \frac{\rho_0x_0 d_0^-(\s_0) + \lambda}{\rho_{t+1}d_0^-(\s_0)}\big)\1_{\{ \rho_t x_t >  \rho_0x_0 +\frac{\lambda}{d_0^-(\s_0)} \} }\label{eqopt_policy_P1}
\end{align}
for $t=0,1,\ldots,T-1$. Under this policy, the expected value and variance of the terminal wealth $x_T^{*}$ are given by
\begin{align}
\begin{dcases}
\E[x_T^{*}] = \rho_0 x_0 +\lambda  {\big(}\frac{1}{d_0^-(\s_0)} -1 \big),\\
\Var[x_T^{*}] =\lambda^2 \big( \frac{1}{d_0^-(\s_0)} -1\big), 
\end{dcases}\label{eqP1_mean_variance}
\end{align}
{respectively}, and they lie on the MV efficient frontier, expressed as,
\begin{align}
\Var[x_T^{*}] = \frac{d_{0}^-(\s_0)  \big(\E[x_T^{*}]-\rho_0x_0 \big)^2}{1-d_{0}^-(\s_0) }\label{eqP1_eff_frontier}
\end{align}
for $\E[x^{*}_T] \geq \rho_0 x_0$.
\end{theorem}

\proof Upon examining the objective functions of problem $\mathcal{L}(\lambda,b)$ and problem $\cP^1_{\mv}(\lambda)$, it becomes evident that if the parameter $b$ satisfies $b=\E_0[x_T^{*}]$, where $x_T^{*}$ represents the terminal wealth generated by implementing the optimal portfolio policy (\ref{eqpolicy_problem_A}), then the objective functions of these two problems are equal. In other words, when $b=\E_0[x_T^{*}]$, the policy (\ref{eqpolicy_problem_A}) also solves problem $\cP^1_{\mv}(\lambda)$. On the other hand, the objective function of $\mathcal{L}(\lambda,b)$ implies that
\begin{align}
\E[x_T] &= \arg \min_b v(\mathcal{L}(\lambda, b)) \notag\\
        &=\arg \min_b \big\{\E_0\big[ (x_T -b)^2\big]  - 2 \lambda  \E[x_T] \big\}. \label{eqP1_obj_b}
\end{align}
That is to say, finding the minimizer of problem (\ref{eqP1_obj_b}) provides a way to identify a proper $b$ that satisfies $b=\E_0[x_T^{*}]$. We then compute the value $v(\mathcal{L}(\lambda, b))$. Notably, the difference between the objective values of problem $\cA(\lambda,b)$ and $\mathcal{L}(\lambda, b)$ is $2\lambda b +\lambda^2$. Using (\ref{def_J_0}) and replacing the variable $y_0$ by (\ref{def_yt_xt}), it has 
\begin{align}
&v(\mathcal{L}(\lambda,b))= \rho_0^2 \big(  d_0^-(\s_0) \big(x_0 -\frac{b +\lambda}{\rho_0} \big)^2
                            \1_{ \big\{ x_0 \leq \frac{b+\lambda}{\rho_0} \big\}} \notag\\
&~~~+ d_0^+(\s_0) \big(x_0 -\frac{b +\lambda}{\rho_0} \big)^2
                            \1_{ \big\{ x_0 > \frac{b+\lambda}{\rho_0} \big\}}\big)-2\lambda b -\lambda^2.\label{eqL_obj}
\end{align}
As the above function has two pieces, we {consider} each piece {separately}. For the piece with $x_0 \leq  \frac{b+\lambda}{\rho_0}$, we denote the minimizer in (\ref{eqP1_obj_b}) as $b^-$. It has
\begin{align}
b^- =&\arg \min_b \big\{ \rho_0^2 d_0^-(\s_0) 
\big( x_0 -\frac{b +\lambda}{\rho_0} \big)^2 - 2\lambda b -\lambda^2 \big\}\notag\\
=&\rho_0 x_0 +\lambda \big( \frac{1}{ d_0^-(\s_0)} -1 \big).\label{eqb_minus}
\end{align}
From Lemma \ref{lem-inequ}, it has $d_0^-(\s_0)>0$. Together with $\lambda\geq 0$, we have $\frac{b^- + \lambda}{\rho_0} = x_0 +\frac{\lambda}{d_0^-(\s_0) \rho_0}\geq x_0$, which means that the value $b^-$ {satisfies} the condition $x_0 < \frac{b+\lambda}{\rho_0}$. Then, substituting $b^-$ in policy (\ref{eqpolicy_problem_A}) and replace the state $y_t$ by (\ref{def_yt_xt}) give the optimal policy (\ref{eqopt_policy_P1}) for problem $\cP^1_{\mv}(\lambda)$. We then consider the other piece with $ x_0 \geq \frac{b + \lambda}{\rho_0}$. we denote the minimizer in (\ref{eqP1_obj_b}) as $b^+$. Using the similar argument, it has  $b^+ $ $=$ $\E[x_T^*] =\rho_0 x_0 +\lambda \big( \frac{1}{ d_0^+(\s_0)} -1 \big)$.
However, it has $\frac{b^+ + \lambda}{\rho_0} = x_0 +\frac{\lambda}{d_0^+(\s_0) \rho_0}\geq x_0$, which does not {satisfy} $ x_0 > \frac{b^- +\lambda}{\rho_0}$. Thus, there does not exit $b^+$ that solves $b^+ =\E[x_T^{*}]$.

We then prove (\ref{eqP1_mean_variance}). Clearly, $\E[x_T^*]$ is given in (\ref{eqb_minus}). We only need to compute $\Var_0[x_T^{*}]$. Using (\ref{eqL_obj}), it has $\Var_0[x_T^{*}] $ $=$ $ v(\mathcal{L}(\lambda,b)) + 2\lambda b^- $ $=$ $\lambda^2 \big( \frac{1}{d_0^-(\s_0)} -1\big)$. At last, we get (\ref{eqP1_eff_frontier}) by eliminating $\lambda$ in (\ref{eqP1_mean_variance}).\endproof

Lemma \ref{lem-inequ} shows that $d_0^-(\s_0)\in (0,1]$. Thus, (\ref{eqP1_mean_variance}) implies that $\E[x_T^{*}]\geq x_0\rho_0$ and $\Var_0[x_T^{*}]\geq 0$ for any $\lambda\geq 0$. 

The alternative formulation $\cP_{\mv}^2(x_{\tg})$ can be solved through the duality relationship with problem $\cP_{\mv}^1(\lambda)$ as follows.
\begin{theorem}\label{thm_policy_P2}
The solution of problem  $\cP_{mv}^2(x_{\tg})$ has the following cases: 
\begin{itemize}
\item[(i)] When $\rho_0 x_0 \leq x_{\tg}$, $d_0^-(\s_0)=1$, or $\rho_0 x_0 \geq x_{\tg}$, $d_0^+(\s_0)=1$, the problem $\cP_{mv}^2(x_{\tg})$ does not have feasible policy. 

\item[(ii)] When $\rho_0 x_0 \leq x_{\tg}$, $d_0^-(\s_0)<1$, the optimal policy of problem $\cP_{\mv}^2(x_{\tg})$ is, 
\begin{align}
&\bm{\pi}^{*}_{t}(x_t,\s_t)=-\bm{k}_t^-(\s_t) \bigg( r_{t+1}^0 x_t - \frac{x_{\tg} - \rho_0  d_0^{-}(\s_0) x_0 }{ \rho_{t+1}(1-d_0^-(\s_0))} \bigg)\notag\\
&\times \1_{ \big\{ \rho_t x_t \leq \frac{ x_{\tg} -\rho_0 d_0^-(\s_0) x_0 }{1- d_0^-(\s_0)} \big\} }
+\bm{k}_t^+(\s_t)  \bigg( r_{t+1}^0 x_t  \notag\\
&- \frac{x_{\tg} - \rho_0  d_0^{-}(\s_0) x_0 }{ \rho_{t+1} (1-d_0^-(\s_0) )}\bigg)
 \1_{ \big\{ \rho_t x_t \leq \frac{ x_{\tg} -\rho_0 d_0^-(\s_0) x_0 }{1- d_0^-(\s_0) } \big\} }\label{eqopt_policy_P2}
\end{align}
By setting $\E[x^{*}] = x_{\tg}$, the MV efficient frontier is given by (\ref{eqP1_eff_frontier}).
\end{itemize}
\end{theorem}
\proof We introduce {Lagrange} multiplier $\lambda\in \mR$ for the constraint $\E[x_T] = x_{\tg}$ in Problem $\cP_{\mv}(x_{\tg})$, which gives the following problem $\mathcal{L}(x_{\tg},\lambda)$,
\begin{align*}
\min_{\{\bm{\pi}_t \in \cA_t\}|_{t=0}^{T-1}}\quad &\Var_0[x_T] - 2\lambda  \big(\E_0[x_T] -x_{\tg}\big) \notag\\
\mbox{s.t.}~~~&~\textrm{$\{x_t, \bm{\pi}_t,\r_t,\s_t \}$ {satisfies} (\ref{eqr}), (\ref{eqs})},(\ref{def_xt}).
\end{align*}
The above problem is similar to problem $\cP_{\mv}(\lambda)$ by adding constant $\lambda x_{\tg}$. The weak duality relationship implies that $v(\mathcal{L}(x_{\tg}, \lambda))\leq v(\cP_{\mv}(x_{\tg}))$. The strong duality holds when the $\E[x_T^{*}]$ generated from problem $\mathcal{L}(x_{\tg},\lambda)$ satisfies the constraint $ \E[x_T^{*}] = x_{\tg}$. Using the result in (\ref{eqP1_mean_variance}), we have
\begin{align}
\E_0[x_T^{*}] = \rho_0 x_0 + \lambda \big(\frac{1}{d_0^-(\s_0)} -1 \big) = x_{\tg}. \label{eqlambda}
\end{align} 
Note that Lemma \ref{lem-inequ} shows that $d_0^-(\s_0)>0$. Clearly, when $x_{\tg}< \rho_0 x_0$ or $\d_0^-(\s_0) =1$, the above equation admits no solution. When $x_{\tg}> \rho_0 x_0$ and $\d_0^-(\s_0) <1$, then equation (\ref{eqlambda}) has the solution
\begin{align}
\lambda^* = \frac{d_0^-(\s_0) (x_{\tg} -\rho_0 x_0)}{ 1-d_0^-(\s_0)}.\label{eqlambda_opt}
\end{align}
Then substituting $\lambda^*$ to the policy (\ref{eqopt_policy_P1}) gives to optimal policy (\ref{eqopt_policy_P2}) for problem $\cP_{\mv}(x_{\tg})$. Varying the target wealth $x_{\tg}$, we may get different $\lambda$. Using (\ref{eqP1_mean_variance}), we may get the MV efficient frontier similar to (\ref{eqP1_eff_frontier}).\endproof

Both optimal portfolio policies (\ref{eqopt_policy_P2}) and (\ref{eqopt_policy_P1}) are piecewise linear functions of the current wealth level $x_t$. This key structure allows us to separate the parameters that depend on market states $\s_t$ from the wealth level $x_t$ during the investment process. In other words, we can use numerical methods to compute (and fit) the key random variables {$\{d_t^-(\s_t)\}|_{t=0}^T$ and $\{d_t^+(\s_t)\}|_{t=0}^T$} for different states $\s_t$ offline, and then apply them online when the true $\s_t$ is observed. Notably, the portfolio policies (\ref{eqopt_policy_P1}) and (\ref{eqopt_policy_P2}) also encompass the simpler case where excess returns in different periods are i.i.d., by ignoring the factor model in (\ref{eqs}).  

The random variables $\{d_t^-(\s_t)\}|_{t=0}^T$ and $\{d_t^+(\s_t)\}|_{t=0}^T$ not only play a key role in computing the optimal policy, but themself also have straightforward financial meaning. Given the states $x_t$ and $\s_t$, we define the time-$t$'s conditional \textit{Sharpe Ratio} of the optimal terminal wealth resulted from policies (\ref{eqopt_policy_P1}) or (\ref{eqopt_policy_P2}) as follows
\begin{align}ƒ
\textrm{SR}_t(x_t, \s_t) \triangleq \frac{ \E[x_T^{*}|\s_t] - \rho_t x_t  }{\sqrt{\Var_t[x_T^{*}|\s_t]} }, ~~t=0,\ldots, T-1,
\label{def_sharpe} 
\end{align}
where the numerator $\E[x_T^{*}|\s_t] - \rho_t x_t $ represents the additional wealth than the sole investment in the risk-free asset. The following result characterizes $\textrm{SR}_t(x_t, \s_t)$ explicitly.

\begin{proposition}\label{prop-sharpe}
Given states variables $\s_t$ and wealth $x_t$, applying the optimal portfolio policy (\ref{eqopt_policy_P1}) or (\ref{eqopt_policy_P2}),  the time-$t$'s conditional Sharpe Ratio of optimal terminal wealth is
\begin{align}
\textrm{SR}_t(x_t, \s_t) =
\begin{dcases}
\sqrt{\frac{1 - d_t^-(\s_t)}{d_t^-(\s_t)}} &\textrm{if}~~\rho_t x_t \leq x_{\tg},\\
\sqrt{\frac{1 - d_t^+(\s_t)}{d_t^+(\s_t)}} &\textrm{if}~~\rho_t x_t > x_{\tg},
\end{dcases} \label{eqsharpe_ratio}
\end{align}
for {$t=0,\ldots, T-1$}.
\end{proposition}
\proof At any time $t=0,\ldots,T-1 $, we consider the two cases, namely, $x_t \leq x_{\tg}$ and $x_t > x_{\tg}$. For the case $x_t \leq x_{\tg}$, we may consider a truncated version of problem $\cP^2_{\mv}(x_{\tg})$ starting from time $t$ to time $T$ which is denoted by $\cP^2_{\mv}(x_{\tg})|_{t}^T$. We then apply the similar solution {approach} for problem $\cP^2_{\mv}(x_{\tg})$ with starting time being $0$ to solve such truncated problem $\cP^2_{\mv}(x_{\tg})|_{t}^T$. Similar to (\ref{eqP1_eff_frontier}) given in Theorem \ref{thm_policy_P2}, we may get the expression of the efficient frontier for the truncated problem $\cP^2_{\mv}(x_{\tg})|_{t}^T$ as follows
\begin{align*}
\Var[x_T^{*} |\s_t] = \frac{d_t^-(\s_t)\big(\E[x_T^{*}|\s_t]-\rho_tx_t \big)^2}{1-d_t^-(\s_t) }~\mbox{for}~\E_t[x_T^*] \geq \rho_t x_t.
\end{align*} 
Rearranging the above expression, we may get
\begin{align*}
\textrm{SR}_t(x_t, \s_t)= \frac{\E[x_T^{*}|\s_t] -\rho_t x_t }{\sqrt{\Var[x_T^{*}|\s_t]}} = \frac{1 - d_t^-(\s_t)}{d_t^-(\s_t)}
\end{align*}
for $x_t \leq x_{\tg}$. We then consider the second case  $x_t > x_{\tg}$. We may still consider a truncated version of problem $\cP^2_{\mv}(x_{\tg})|_{t}^T$ starting from time $t$ to time $T$. Recall that problem $\cP^2_{\mv}(x_{\tg})$ is solved by using the Lagrange method, i.e., solve the problem $\cP^1_{\mv}(\lambda)$ for some $\lambda$. However, in this case, as $x_t > x_{\tg}$, we need to a problem $\cP^1_{\mv}(\lambda)$ with $\lambda\leq 0$. As the solution approach is similar to the proof of Theorem \ref{thm_policy_P1}, we omit the detail. Then, the MV efficient frontier can be written as
\begin{align*}
\Var[ x_T^{*}|\s_t ] = \frac{d_t^+(\s_t)\big(\E[x_T^{*}|\s_t]-\rho_t x_t \big)^2}{1-d_t^-(\s_t)}~\mbox{for}~\E_t[x_T^{*}] < \rho_t x_t,
\end{align*}  
which further gives the expression of $\textrm{SR}_t(x_t, \s_t)$ as in (\ref{eqsharpe_ratio}).
\endproof
Since the conditional Sharpe Ratio $\textrm{SR}_t(x_t, \s_t)$ is a function of the state variable $\s_t$, it is a random variable influenced by market conditions. Furthermore, (\ref{eqsharpe_ratio}) implies that $\textrm{SR}_t(x_t, \s_t)$ is a decreasing function of the variable $d_t^-(\s_t)$ or $d_t^+(\s_t)$. Therefore, we can consider $d_t^-(\s_t)$ and $d_t^+(\s_t)$ as an evaluation of future investment opportunities (FIO). The smaller the values of $d_t^-(\s_t)$ and $d_t^+(\s_t)$, the better performance can be achieved by the optimal policy in the future periods given current market condition.

\subsection{Numerical Methods}\label{sse_numerical_method}
In the previous section, we derived the semi-analytical optimal policy in Theorems \ref{thm_policy_P1} and \ref{thm_policy_P2}. However, numerical methods are still needed to compute key parameters offline for different market conditions. In this section, we propose algorithms to compute $\k_t^-(\s_t)$, $\k_t^+(\s_t)$, $d_t^-(\s_t)$, and $d_t^+(\s_t)$, which are crucial for characterizing the MMV policies.

We first consider the Markov chain-based regime switching model described in Section \ref{sec_model}. In this setting, the market condition $s_t \in \{1, \ldots, M\}$ for $t=0,\ldots,T$ evolves according to a Markov chain model, and the excess returns $\{\r_t\}|_{t=1}^T$ follow the model in (\ref{def_rt_markov}). The following Algorithm \ref{alg-markov} is suitable for computing the portfolio policy in this setting. Note that, in Algorithm \ref{alg-markov}, although we assume $\{\r_t\}|_{t=1}^T$ follow multivariate normal distributions with parameters controlled by a Markov Chain, this setting can be extended to other distributions as long as their parameters can be characterized by a Markov Chain.    

\begin{algorithm}
\small
\caption{\small Computation of portfolio policy for Markov Chain-based regime switching model}\label{alg-markov}
\begin{algorithmic}[1]
\REQUIRE The parameters $\bm{c}(s)$ and $\bm{\Sigma}(s)$ for all $s \in \{1,2,\ldots, M\}$ defined in (\ref{def_rt_markov}); Transition probability $P_{i,j}$, $i,j \in \{1,\ldots, M\}$.
\ENSURE $\k_t^-(s)$, $\k_t^+(s)$, $d_t^-(s)$, $d_t^+(s)$ for $s \in \{1,\ldots, M\}$ and $t=0,\ldots, T-1$.

\STATE \textbf{Generating samples:} For each possible value of state $j \in \{1,\ldots, M\}$, generate $L$ samples denoted by $\{\r^{(j,\ell)}\}_{\ell=1}^L$ according to the distribution $\mathcal{N}(\bm{c}(j), \bm{\Sigma}(j))$
\STATE Set $d_T^- \leftarrow 1$, $d_T^+ \leftarrow 1$, $t \leftarrow T-1$
\WHILE{$t>0$}
    \FOR{each $s \in \{1,\ldots, M\}$}
        \STATE Compute $\k_{t}^-(s)$ and $d_t^-(s)$ by solving the optimization problem:
        \STATE $d_t^-(s) = \min_{\k \in \mathcal{A}_t} \frac{1}{L} \sum_{j=1}^{M} \sum_{\ell=1}^{L} P_{s, j} \big( 1- \k^{\top} \r^{(j,\ell)} \big)^2
        \times \big( d_{t+1}^-(j) \1_{\{\k^{\top} \r^{(j,\ell)} \leq 1\}} + d_{t+1}^+(j) \1_{\{\k^{\top} \r^{(j,\ell)} > 1\}} \big),$
        \STATE where the solution is $\k_{t}^-(s)$.
        
        \STATE Compute $\k_{t}^+(s)$ and $d_t^+(s)$ by solving the optimization problem:
        \STATE $d_t^+(s) = \min_{\k \in \mathcal{A}_t} \frac{1}{L} \sum_{j=1}^{M}\sum_{\ell=1}^{L} P_{s,j} \big( 1 + \k^{\top} \r^{(j,\ell)} \big)^2  \times \big( d_{t+1}^-(j) \1_{\{\k^{\top} \r^{(j,\ell)} \geq -1\}} + d_{t+1}^+(j) \1_{\{\k^{\top} \r^{(j,\ell)} < -1\}} \big),$
        \STATE where the solution is $\k_{t}^+(s)$.
    \ENDFOR
    \STATE $t \leftarrow t-1$
\ENDWHILE
\end{algorithmic}
\end{algorithm}
\normalsize

We then present an example to demonstrate the procedure outlined in Algorithm \ref{alg-markov}, using the Markov regime-switching market model from \cite{costa2008generalized}. Specifically, we consider four stocks (IBM, Altria Group, Exxon Mobil, UTC) from the Dow Jones Index, with returns differentiated by two market states, $\s_t \in \{S^1, S^2\}$. The model parameters are calibrated following \cite{costa2008generalized}, where the quarterly risk-free return is $r_f = 1.003$. The mean excess returns for these assets are $\bm{c}(S^1) = \big(0.167, 0.157, 0.057, 0.147 \big)^{\top}$ and $\bm{c}(S^2) = \big(-0.193, -0.063, -0.073, -0.113 \big)^{\top}$ for the two states. The covariance matrices for $S^1$ and $S^2$ are: $\bm{\Sigma}(S^1)$$=$$10^{-2}\cdot\begin{pmatrix}
 3.06 & 0.12 & 0.15  &  0.47 \\
 0.12 & 3.19 & 0.32  &  0.27 \\
 0.15 & 0.32 & 1.30  &  0.41 \\
 0.47 & 0.27 & 0.41  &  2.22 \\
\end{pmatrix}$ and $\bm{\Sigma}(S^2)$$=$$10^{-2}
\begin{pmatrix}
4.88 & 0.36 & 1.16  &  1.94 \\
0.36 & 3.69 & 0.69  &  0.64 \\
1.16 & 0.69 & 2.57  &  1.41 \\
1.94 & 0.64 & 1.41  &  5.80 \\
\end{pmatrix}$.
For each state $s_t \in \{S^1, S^2\}$, returns $\r_t \sim \mathcal{N}(\bm{c}(s_t), \bm{\Sigma}(s_t))$. We label $S^1$ as the ``good" market state and $S^2$ as the ``bad" market state. The transition probabilities between states are $\mP(s_{t+1} = S^1 | s_t = S^1) = 0.7$ and $\mP(s_{t+1} = S^2 | s_t = S^2) = 0.6$. We consider the portfolio model $\cP_{\mv}^2(x_{\tg})$ with $T=12$, including both the no-short-selling and the cardinality constraints, i.e., $\mathcal{A}_t $$=$$\{ \bm{\pi} \in \mathbb{R}^n | \bm{\pi} \geq 0, ~~\sum_{i=1}^n |\text{sign}(\pi_i)| < 2 \}$ for $t=0,\ldots, T-1$. Other parameters are $x_0=1$ and $x_{\tg}=1.178$. To apply Algorithm \ref{alg-markov}, we simulate $L=10,000$ samples of returns for each market state $s_t \in \{S^1, S^2\}$. To handle the cardinality constraint, we enumerate all possible combinations of selecting two active assets for the portfolio. For each active portfolio, we use Matlab's optimization toolbox to solve the associated convex optimization problem.

Table \ref{Table_markov} provides detailed values for $d_t^-(s_t)$, $d_t^+(s_t)$, $\k_t^-(s_t)$, and $\k_t^+(s_t)$ for the selected time period. As a reminder, considering the portfolio policy in (\ref{eqopt_policy_P1}), {$\k_t^-(s_t)$ and $\k_t^+(s_t)$} regulate the allocation proportion in two cases when the condition $\rho_t x_t \geq \rho_0x_0 +\lambda/d_0^-(\s_0)$ is met or not. Table \ref{Table_markov} reveals that, due to the cardinality constraint, the active assets in the portfolio differ for {$\k_t^-(s_t)$ and $\k_t^+(s_t)$} (specifically, stock 2 and stock 4 are active for {$\k_t^-(s_t)$}, while stock 1 and stock 3 are active for {$\k_t^+(s_t)$}), regardless of the market states. 

\begin{table}[h]
\centering
\caption{\small The values of $d_t^-(s_t)$, $d_t^+(s_t)$, $\k_t^-(s_t)$, and $\k_t^+(s_t)$ for Markov regime switching model}\label{Table_markov}
\footnotesize
\setlength{\tabcolsep}{3pt}
\begin{tabular}{cc cccc cccc cc}
\toprule
     &       & \multicolumn{4}{c}{$\k_t^-$} & \multicolumn{4}{c}{$\k_t^+$} &  \multicolumn{2}{c}{FIO} \\
              \cmidrule(lr){3-6}   \cmidrule(lr){7-10} \cmidrule(lr){11-12} 
$t$  & state & $k_{t,1}^-$ & $k_{t,2}^-$ & $k_{t,3}^-$ & $k_{t,4}^-$ 
             & $k_{t,1}^+$ & $k_{t,2}^+$ & $k_{t,3}^+$ & $k_{t,4}^+$ & $d_t^-$ & $d_t^+$ \\
\cmidrule(lr){1-2}     \cmidrule(lr){3-6}   \cmidrule(lr){7-10}         \cmidrule(lr){11-12} 
\multirow{2}{*}{$0$} & $S^1$ & 0.00   & 1.33    & 0.00    & 0.55    & 0.00    & 0.00    & 0.00    & 0.00    & 0.32    & 0.38\\
                     & $S^2$ & 0.00   & 0.34    & 0.00    & 0.00    & 0.71    & 0.00    & 0.53    & 0.00    & 0.40    & 0.38\\
\cmidrule(lr){1-2}     \cmidrule(lr){3-6}   \cmidrule(lr){7-10} \cmidrule(lr){11-12}
\multirow{2}{*}{$1$} & $S^1$ & 0.00   & 1.33    & 0.00    & 0.55    & 0.00    & 0.00    & 0.00    & 0.00    & 0.35    & 0.41\\
                     & $S^2$ & 0.00   & 0.34    & 0.00    & 0.00    & 0.71    & 0.00    & 0.53    & 0.00    & 0.43    & 0.41\\
\cmidrule(lr){1-2}     \cmidrule(lr){3-6}   \cmidrule(lr){7-10} \cmidrule(lr){11-12}
$\vdots$ & $\vdots$ & $\vdots$ & $\vdots$ & $\vdots$ & $\vdots$ & $\vdots$ & $\vdots$ & $\vdots$ & $\vdots$ & $\vdots$ \\
\cmidrule(lr){1-2}     \cmidrule(lr){3-6}   \cmidrule(lr){7-10} \cmidrule(lr){11-12}
\multirow{2}{*}{$11$}& $S^1$ & 0.00    & 1.42    & 0.00    & 0.74    & 0.00    & 0.00    & 0.00    & 0.00    & 0.82    & 0.99\\
                     & $S^2$ & 0.00    & 0.56    & 0.00    & 0.00    & 0.51    & 0.00    & 0.48    & 0.00    & 0.99    & 0.97\\
\bottomrule
\end{tabular}
\end{table}

\begin{figure}[htp!]
\caption{\small Comparison of portfolio and efficient frontiers under Markov regime switching model {(add centering to subfigures)}}
\centering
\begin{subfigure}{0.4\textwidth}
    \centering
    \includegraphics[width = 0.8\linewidth]{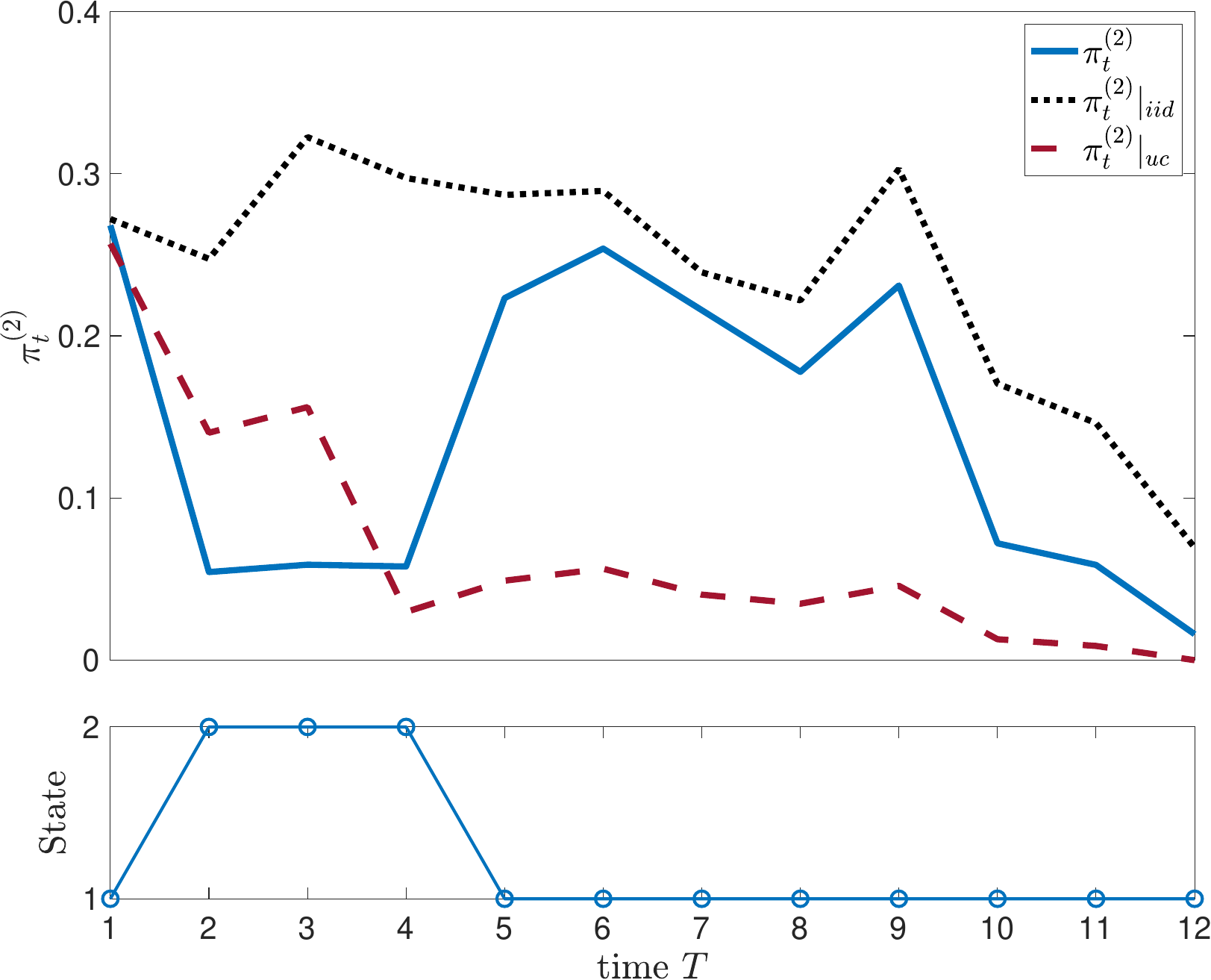}
    \caption{\footnotesize The portfolios}
    \label{figexam_markov_portfolio}
\end{subfigure}
\begin{subfigure}{0.4\textwidth}
    \centering
    \includegraphics[width = 0.8\linewidth]{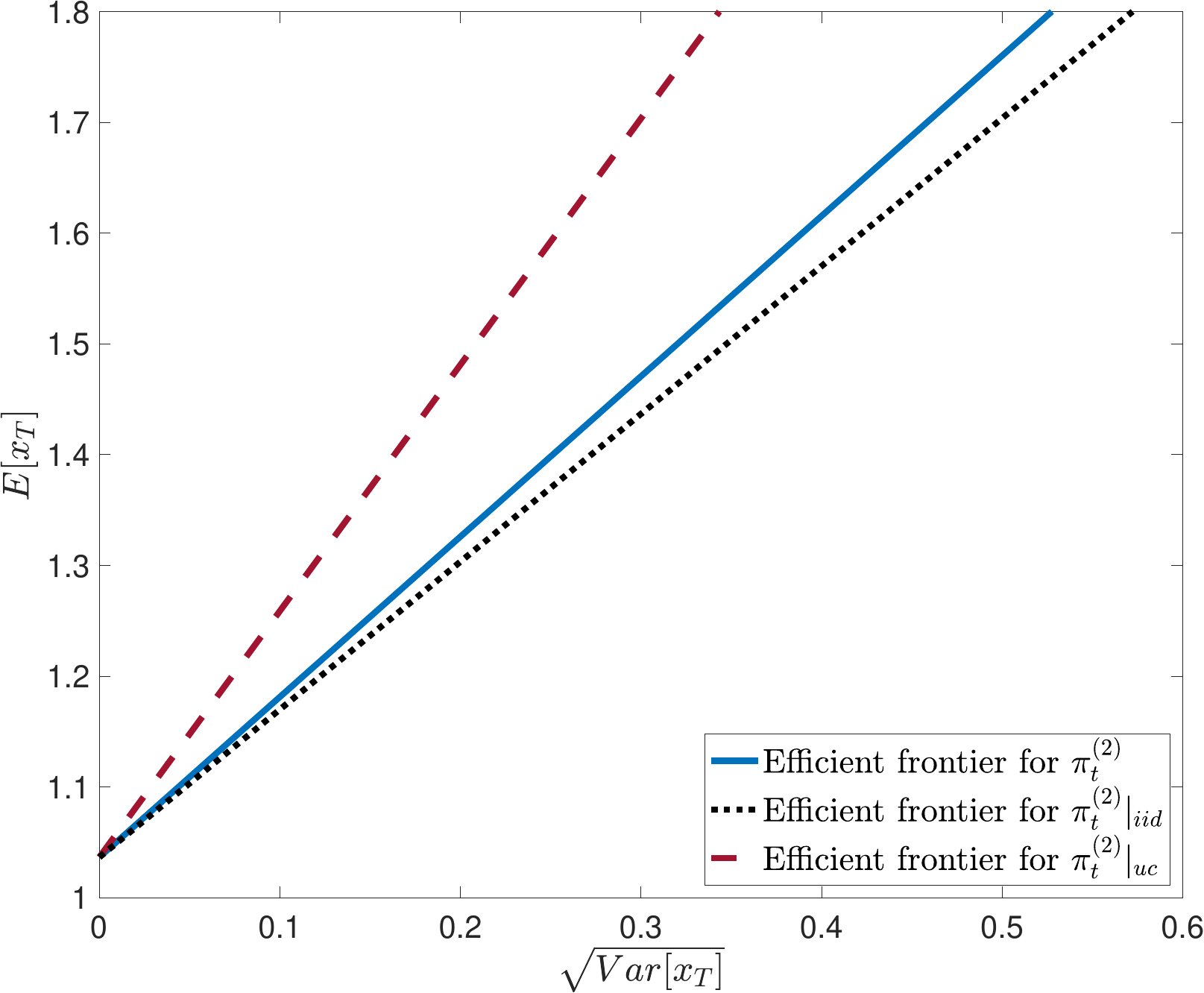}
    \caption{\footnotesize The MV efficient frontiers}
    \label{figexam_markov_efficient}
\end{subfigure}\label{figexam_markov}
\end{figure}

Figure \ref{figexam_markov_portfolio} illustrates sample paths of the portfolio holdings for the second asset, generated by different models in response to a single sample path of risky asset prices. The corresponding market state path, $s_t$, is also shown. Specifically, the portfolio from our model, $\cP_{\mv}(x_{\tg})$ is represented by the solid blue line (denoted as {$\bm{\pi}^{(2)}_t$}). We also compare two benchmark models. The first, represented by the dotted black line (denoted as {$\bm{\pi}^{(2)}_t|_{iid}$}), uses the same portfolio constraint but ignores return predictability by assuming the market state $s_t$ is independent over time and follows the stationary distribution ($\mP(s_t=S^1)$$=$$0.571$ and $\mP(s_t=S^2)$$=$$0.429$). The second benchmark model, denoted by {$\bm{\pi}_t^{(2)}|_{uc}$} (shown by the red dashed line), retains return predictability but disregards the portfolio constraints.

It is evident from Figure \ref{figexam_markov_portfolio} that these portfolios exhibit different patterns in response to the identical asset price path. The cardinality-constrained portfolio {$\bm{\pi}_t^{(2)}$} accounts for return predictability, and its weights are more concentrated on a small group of assets, making it more sensitive to state variations compared to the other two portfolios. In periods $t=6,\ldots,12$, as the state remains in $s_t=S^1$, portfolios {$\bm{\pi}_t^{(2)}$ and $\bm{\pi}_t^{(2)}|_{iid}$} exhibit similar patterns. However, the unconstrained portfolio {$\bm{\pi}_t^{(2)}|_{uc}$} behaves dramatically differently from the other two portfolios. Integrating the Markov regime switching enables to track the market changes more actively than the other portfolio.  Figure \ref{figexam_markov_efficient} further compares the MV efficient frontiers resulting from these portfolios. We observe that, as portfolio {$\bm{\pi}_t^{(2)}$}  capitalizes on Markov regime-switching returns, its efficient frontier has a higher slope (higher Sharpe Ratio) than the one generated by {$\bm{\pi}_t^{(2)}|_{iid}$}, which ignores the regime-switching structure by assuming independent returns. On the other hand, in the ideal situation, the unconstrained portfolio {$\bm{\pi}_t^{(2)}|_{uc}$} generates a higher Sharpe ratio than the one from the constrained portfolio {$\bm{\pi}_t^{(2)}$}. In summary, this comparison indicates that portfolio constraints and return predictability significantly impact MV portfolio holdings and overall performance. It is essential to be cautious about oversimplifying models by either ignoring returns' predictability or portfolio constraints.

We proceed to outline the numerical solution process for dynamic factor models. In contrast to Markov chain-based models, which inherently possess a discrete state space, dynamic factor models are typically characterized by continuous space, i.e., $\s_t\in \mR^{N_s}$. Consequently, we need to discretize the state space and calculate the associated variables. Once these discrete points are computed, an interpolation method is applied to recover variables such as $d_t^{-}(\s_t)$, $d_t^{-}(\s_t)$, $\k_t^{-}(\s_t)$, and $\k_t^{+}(\s_t)$ in continuous state space. The detailed procedure is outlined in Algorithm \ref{alg-factor} below.

\begin{algorithm}
\small
\caption{\small Computation of portfolio policy for the dynamic factor model}\label{alg-factor}
\begin{algorithmic}[1]
\REQUIRE The parameters in $\mathbf{f}(t,\s_t,\epsilon_t)$ and $\mathbf{g}(t,\s_t, \xi_t)$, $t=0,\ldots, T-1$; number of return samples $L$ for each state realization;
\ENSURE $\k_t^-(\s)$, $\k_t^+(\s)$, $d_t^-(\s)$, $d_t^+(\s)$ for $\s \in \mathbb{R}^{N_s}$ and $t=0,\ldots, T-1$
\STATE Simulate the finite discrete set of states as $\{\s_{t}^{(j)}\}$ for $j=1,\ldots, J$
\FOR{$j=1$ \TO $J$}
    \FOR{$\ell=1$ \TO $L$}
        \STATE Simulate $(\s_{t+1}^{(j,\ell)},\r_{t+1}^{(j,\ell)})$ based on $\s_{t+1} = \mathbf{g}(t+1,\s_{t},\bm{\xi}_{t+1})$ and $\r_{t+1} = \mathbf{f}(t+1, \s_{t+1},\bm{\epsilon}_{t+1})$
    \ENDFOR
\ENDFOR
\STATE $d_T^-(\s_T) \leftarrow 1$, $d_T^+(\s_T) \leftarrow 1$ for all $\s_T\in \mathbb{R}^m$
\STATE $t \leftarrow T-1$
\WHILE{$t>0$}
    \FOR{$j=1$ \TO $J$}
        \STATE Compute $\k_t^-(\s_{t}^{(j)})$ by solving the optimization problem:
        \begin{align*}
        &d_t^-(\s_t^{(j)}) \leftarrow \min_{\k \in \mathcal{A}_t} \frac{1}{L} \sum_{\ell=1}^{L} \big( 1-\k^{\top} \r_{t+1}^{(j,\ell)} \big)^2  \Big(d_{t+1}^-(\s_{t+1}^{(j,\ell)}) \\
        &\times\1_{\{\k^{\top} \r_{t+1}^{(j,\ell)} \leq 1\}} + d_{t+1}^+(\s_{t+1}^{(j,\ell)})\1_{\{ \k^{\top} \r_{t+1}^{(j,\ell)} >1\}}\Big);
        \end{align*}
        \STATE Compute $\k_{t}^+(\s_{t}^{(j)})$ by solving the optimization problem:
        \begin{align*}
        &d_t^+(\s_t^{(j)}) \leftarrow \min_{\k\in \mathcal{A}_{t}} \frac{1}{L} \sum_{\ell=1}^{L}\big(1+\k^{\top} \r_{t+1}^{(j,\ell)}\big)^2  \Big( d_{t+1}^-(\s_{t+1}^{(j,\ell)}) \\
        &\times \1_{\{\k^{\top} \r_{t+1}^{(j,\ell)}\geq 1 \}} 
        + d_{t+1}^+(\s_{t+1}^{(j,\ell)}) \1_{\{\k^{\top} \r_{t+1}^{(j,\ell)}<1\}}\Big);
        \end{align*}
    \ENDFOR
    \STATE Based on the data $d_t^-(\s_t^{(j)})$, $d_t^+(\s_t^{(j)})$, $\k_t^-(\s_{t}^{(j)})$ and $\k_t^+(\s_{t}^{(j)})$, $j=1,\ldots, J$, approximate the functions $d_{t}^-(\s_t)$, $d_{t}^+(\s_t)$, $\k_{t-1}^-(\s_t)$ and $\k_{t-1}^+(\s_t)$ with respect to $\s_t$ by curve fitting.
    \STATE $t \leftarrow t-1$
\ENDWHILE
\end{algorithmic}
\end{algorithm}
\normalsize

Algorithm \ref{alg-factor}, which discretizes the state space, is best suited for low-dimensional models like the three- or five-factor models in \cite{fama1993common,fama2015five,garleanu2013dynamic}. A key step involves approximating the functions $\k_t^-(\s_t)$, $\k_t^+(\s_t)$, $d_t^-(\s_t)$, and $d_t^+(\s_t)$, for which any curve-fitting method can be used (see \cite{Zielesny-book-curve-fit}). However, Algorithm \ref{alg-factor} may cause an exponential increase in grid realizations as state dimensions grow, particularly for states driven by multivariate normal distributions, where extreme values are rare. To address this in high-dimensional spaces, we propose two modifications: (1) use random sampling to generate ${\s_{t}^{(j)}}|_{j=1}^J$, reducing low-probability samples; and (2) employ \textit{neural networks} for function approximation, which handle uneven grid distributions more effectively (see \cite{andras2017high,Zielesny-book-curve-fit}). Our numerical tests (Section \ref{sec_empirical_analysis}) show this method performs stably.

\section{Time Consistency in Efficiency of the Optimal Policy}\label{sectime-consistency}
The optimal portfolio policies (\ref{eqopt_policy_P1}) and (\ref{eqopt_policy_P2}) derived for problems $\cP_{\mv}^1(\lambda)$ and $\cP_{\mv}^2(x_{\tg})$ are known as the \textit{pre-committed policy} (see \cite{basak2010dynamic,cui2012better}). This type of policy is not time consistent, meaning the policy derived at time $t=0$ is not consistent with the policy derived at any intermediate time point $0<t<T$. This inconsistency may lead to irrational behaviors during investment when applying the optimal policy (see \cite{cui2017mean}). As a remedy for the pre-committed policy, \cite{cui2012better} and \cite{cui2017mean} propose a concept called \textit{Time Consistency in Efficiency} (TCIE). This concept means that the policy derived from the MV objective function $\Var[x_T] - 2\lambda \E[x_T]$ remains optimal for the truncated problem at time $0<t<T$ with the objective function $\Var[x_T|\cF_t] - 2\lambda \E[x_T|\cF_t]$ for some $\lambda > 0$. If $\lambda$ exists, the truncated pre-committed policy may still result in the MV efficient terminal wealth, albeit at a different position on the MV efficient frontier. Therefore, the crucial question is to develop the condition to guarantee the existence of such an MV tradeoff parameter $\lambda$. \cite{cui2017mean} reveals that this condition is related to the variance-optimal signed supermartingale measure (VSSM) and develop sufficient conditions. This section further extends this result to our dynamic factor-based MMV model. We first characterize the density of VSSM using the FIO and then develop the condition for VSSM such that the pre-committed optimal policies, i.e. (\ref{eqopt_policy_P1}) and (\ref{eqopt_policy_P2}), satisfy TCIE property. In the following part, we use $\mathcal{L}^2 (\mathcal{F}_{t},\textit{P}) $ to denote the set of all $\mathcal{F}_{t}$-measurable square integrable random variables.

\begin{definition}
    A signed measure $\mQ$ on the filtered space $\{\Omega, \{\mathcal{F}_t\}|_{t=0}^{T}, \mP\}$ is called a signed supermartingale measure, if $\mQ(\Omega)=1$, $\mQ\ll \mP$ in $\mathcal{L}^2 (\mathcal{F}_{T},\mP)$ and the discounted wealth process of any admissible policy is a supermartingale under $\mQ$, i.e., for $t = 0,1,\cdots, T-1$,
    \begin{align}
    &\E \Big[ \frac{\mathrm{d}\mQ}{\mathrm{d} \mP} \rho_t^{-1} x_T(\bm{\pi}_0, \bm{\pi}_1, \cdots \bm{\pi}_{T-1}) \mid \mathcal{F}_t \Big] \notag \\
    &~~~~~~\leq  x_t(\bm{\pi}_0, \bm{\pi}_1, \cdots \bm{\pi}_{t-1}),\quad \forall \bm\pi_t \in\cA_t, \label{defmar}
    \end{align}
    where $x_t(\bm{\pi}_0, \bm{\pi}_1, \cdots \bm{\pi}_{t-1})$ denotes the wealth at time $t$ by applying policy $\{\bm{\pi}_0, \bm{\pi}_1, \cdots \bm{\pi}_{t-1}\}$, $\frac{\mathrm{d}\mQ}{\mathrm{d} \mP}$ denotes the density of the signed measure $\mQ$ with respect to probability $\mP$.
\end{definition}

\begin{definition}
    Denote $\mathcal{P}_s$ as the set of all the signed supermartingale measures. A measure $\tilde{\mP}\in \cP$ is called variance-optimal signed supermartingale measure (VSSM), if $\tilde{\mP}$ has the smallest variance in $\mathcal{P}_s$, i.e.,
    \begin{align}
    \tilde{\mP} = \arg\min_{\mQ \in \mathcal{P}_s}~~\Var \Big[\frac{\mathrm{d}\mQ}{\mathrm{d} \mP}\Big] .\label{eqVSMM}
    \end{align}
\end{definition}

We then compute the VSSM of our market. For any signed measure $\mQ$, by defining
\begin{align*}
m_t \triangleq \E \Big[ \frac{\mathrm{d}\mQ}{\mathrm{d} \mP}  \Big | \mathcal{F}_t \Big] \Big/ \E \Big[ \frac{\mathrm{d}\mQ}{\mathrm{d} \mP} 
\Big| \mathcal{F}_{t-1} \Big],~~t=1,\ldots, T,
\end{align*}
it has $\frac{\mathrm{d}\mQ}{\mathrm{d} \mP} = m_1  m_2 \cdots m_T$. Due to the linear relationship between the wealth and the policy, the inequality $(\ref{defmar})$ is equivalent to, 
\begin{align*}
&\E \Big[ \frac{\mathrm{d} \mQ}{\mathrm{d} \mP} \r^{\top}_{t+1} \bm{\pi}_t\Big | \mathcal{F}_t \Big] \leq 0,~~\forall \bm \pi_t \in \cA_t,~t=0,1,\cdots, T-1,
\end{align*}
which further implies $\E \big[ \frac{\mathrm{d} \mQ}{\mathrm{d} \mP} \r_{t+1} \Big | \mathcal{F}_t \big] \in \cA_t^{^\perp}$, $t=0,1,\cdots, T-1$, where  $\cA_t^{\perp}$$=$$\{\bm y \in\mathbb{R}^N ~|~ \bm y^{\top} \bm \pi \leq 0, \bm \pi \in \cA_t\}$ representing the polar cone of $\cA_t$. Then the VSSM of the market can be characterized by the following problem,
\begin{align*}
(\mathcal{P}_{\VSSM})~\min_{m_1,\ldots, m_{T}} \quad &\E[ m_1^2 m_2^2 \cdots  m_{T}^2 ],\\
\mbox{s.t.}\quad &\E[m_{t+1}| \mathcal{F}_t] = 1,\\
&\E[ m_1m_2 \cdots  m_{T}\r_{t+1}| \mathcal{F}_t] \in \cA_t^{^\perp}, \\
&m_{t+1} \in \mathcal{L}^2 (\mathcal{F}_{t+1},\mP),~t = 0,\ldots, T-1.
\end{align*}
In the above formulation, the objective function is obtained based on the following fact:
\begin{align*}
\Var\Big[ \frac{\mathrm{d}\mQ}{\mathrm{d} \mP} \Big]= \E \Big[ \Big( \frac{\mathrm{d}\mQ}{\mathrm{d} \mP} -1 \Big)^2 \Big] = \E \Big[ \Big( \frac{\mathrm{d}\mQ}{\mathrm{d} \mP}  \Big)^2  \Big] -1.
\end{align*}
Problem $(\mathcal{P}_{\VSSM})$ can be solved by dynamic programming which leads to the following result.
\begin{theorem}\label{thm-VSSM}
    The density of VSMM can be expressed as
    \begin{align}
    \frac{\mathrm{d} \tilde{\mP}}{\mathrm{d} \mP} = \frac{1}{ d_0^-(\s_0)} \prod_{i=0}^{T-1}B_i(\s_i), \label{eq_VSMM_desity}
    \end{align}
    where $B_0(\s_0) = 1-\r_1^{\top} \k_0^-(\s_0)$ and $B_t(\s_t)=\big( \big(1-\r_{t+1}^{\top} \k_t^-(\s_t)\big) \1_{\{\prod_{i=0}^{t-1}B_i(\s_i)\geq 0\}}+\big(1+\r_{t+1}^\top \k_t^+(\s_t) \big)\cdot \1_{\{\prod_{i=0}^{t-1}B_i(\s_i)< 0\}}\big)$,
    for $t=1,\ldots,T-1$. Moreover, the conditional expectation of $\frac{\mathrm{d} \tilde{\mP}}{\mathrm{d} \mP}$ is given as 
\begin{align}\label{eqvssm_exp}
&\E\Big[\frac{\mathrm{d} \tilde{\mP}}{\mathrm{d} \mP}\Big|\mathcal{F}_t\Big] 
= \frac{1}{ d_0^-(\s_0)}  \prod_{i=0}^{t-1} B_i(\s_i)\cdot \big(d_t^-(\s_t)\cdot \1_{\{m_1^*m_2^*\cdots
    m_{t}^*\geq  0\}}\notag\\
    &~~~~~~~~~~~+ d_t^+(\s_t)\cdot \1_{\{m_1^*m_2^*\cdots m_{t}^*<  0\}}\big).
\end{align}
\end{theorem}
\proof By using the induction method, we prove that the value function of problem $\mathcal{P}_{\VSSM}$ at time $t$ is in the following {form},
\begin{align}
&J_t(m_1,\dots, m_t,\s_t)=\frac{1}{d_t^-(\s_t)} m_1^2m_2^2
m_{t}^2\cdot \1_{\{m_1m_2\cdots m_{t}\geq 0\}}\notag\\
&~~+\frac{1}{d_t^+(\s_t)}m_1^2m_2^2\cdots m_{t}^2 \cdot \1_{\{m_1m_2\cdots
    m_{t}< 0\}}.\label{claim_vssm}
\end{align}
At time $T$, the claim (\ref{claim_vssm}) holds true automatically as $d_T^-(\s_T)=d_T^+(\s_T)=1$. We assume that the statement (\ref{claim_vssm}) holds true for time $t+1$. We then prove that the claim  (\ref{claim_vssm}) is also true for time $t$. At time $t$, when $m_1m_2\cdots m_{t}> 0$, the Bellman equation can be written as
\begin{align*}
&J_t(m_1,\ldots, m_t,\s_t) \\
=& \min_{ m_{t+1}\in\mathcal{L}^2(\mathcal{F}_{t+1},\mP) }~m_1^2 \cdots m_t^2\E\big[\big(\displaystyle\frac{1}{d_{t+1}^-(\s_{t+1})}\1_{\{m_{t+1}\geq  0\}}
\\
&\quad +\frac{1}{d_{t+1}^+(\s_{t+1})} \1_{\{m_{t+1}< 0\}}\big)m_{t+1}^2\big\vert\s_{t} \big].\\
&\quad \textrm{s.t.}~~\E[m_{t+1}|\mathcal{F}_{t} ] = 1,~\E\big[m_{t+1}\r_{t+1}\big|\mathcal{F}_{t}\big]\in\cA_{t}^\perp,  
\end{align*}
By omitting the term $m_1^2m_2^2\cdots m_{t}^2$, the above problem can be further reduced into the following form,
\begin{align}
\cA_{\VSSM,t}^+~\min_{m_{t+1}\in\mathcal{L}^2(\mathcal{F}_{t+1},\mP) }&~~\E\big[\big(\displaystyle\frac{1}{d_{t+1}^-(\s_{t+1})} \1_{\{m_{t+1}\geq  0\}}\notag\\
       ~&+ \frac{1}{d_{t+1}^+(\s_{t+1})} \1_{\{m_{t+1}< 0\}}\big)m_{t+1}^2 \big\vert\s_{t} \big]. \notag\\
\textrm{s.t.}~~~ &~\E[m_{t+1}|\mathcal{F}_{t} ] = 1,\label{VSSM_eq1}\\
             ~&~ \E\big[m_{t+1}\r_{t+1}\big|\mathcal{F}_{t}\big]\in\cA_{t}^\perp. \label{VSSM_eq2}
\end{align}
Introduce the {Lagrange} multipliers, $\nu\in \mR$ and $\bm{\lambda} \in \mR^N$ for the constraints (\ref{VSSM_eq1}) and (\ref{VSSM_eq2}), respectively, we get the dual problem of $\cA_{\VSSM,t}^+$ with respect to $m_{t+1}\in\mathcal{L}^2(\mathcal{F}_{t+1})$:
\begin{align}
\max_{\nu\in\mathbb{R},~-\bm \lambda\in\cA_{t}}~\mathcal{D}_t(\nu,\lambda)\triangleq \E\big[\min_{m_{t+1} }L_t(m_{t+1},\nu,\bm \lambda)\Big|\s_{t}
\big], \label{VSSM_dual}
\end{align}
where the Lagrangian function is defined as
\begin{align*}
&L_t(m_{t+1},\nu,\bm \lambda)\\
=&\Big(\frac{1}{d_{t+1}^-(\s_{t+1})}\1_{\{m_{t+1}\geq
    0\}}+\frac{1}{d_{t+1}^+(\s_{t+1})} \1_{\{m_{t+1}< 0\}}\Big)
m_{t+1}^2\\
&-\nu ( m_{t+1}-1)-\bm\lambda^{\top}\r_{t+1}m_{t+1}.
\end{align*}
Checking the optimality condition of inner optimization in (\ref{VSSM_dual}) gives
\begin{align}
&m_{t+1}=\frac{d_{t+1}^-(\s_{t+1})}{2}(\nu+\bm \lambda^{\top}\r_{t+1})\1_{\{\nu+\bm \lambda^{\top}\r_{t+1}\geq 
    0\}}\notag\\
&\quad ~~~~~+\frac{d_{t+1}^+(\s_{t+1})}{2}(\nu+\bm \lambda^{\top}\r_{t+1}) \1_{\{\nu+\bm \lambda^{\top}\r_{t+1}< 
    0\}}.\label{eqnm}
\end{align}
Note that $m_{t+1} \geq  0$ if and only if
$\nu+\bm \lambda^{\top}\r_{t+1} \geq  0$. Then, we have
\begin{align*}
&\mathcal{D}_t(\nu,\lambda)\\
=&\E\Big[-\frac{1}{4}(\nu+\bm \lambda^{\top}\r_{t+1})^2 \big( d_{t+1}^-(\s_{t+1})\1_{\{\nu+\bm \lambda^{\top}\r_{t+1}\geq 0\}}\\
&+ d_{t+1}^+(\s_{t+1})\1_{\{\nu+\bm \lambda^{\top}\r_{t+1}<0\}} \big) +\nu ~\Big|\s_t\Big].
\end{align*}
If $\nu>0$, identifying the optimal $\bm \lambda$ within the cone $-\bm \lambda \in \cA_t$ is equivalent to identifying the optimal $\k_t$ within the cone $\k_t \in \cA_t$ when we set $\bm \lambda=- \nu \k_t$. Then, it has
\begin{align*}
&\max_{\nu>0, -\bm \lambda\in\cA_{t}}\mathcal{D}_t(\lambda,\nu) \\
=&\max_{\nu>0, \k_t \in\cA_{t}} \E\Big[ -\frac{1}{4} \nu^2 (1-\k_t^{\top}\r_{t+1})^2\big( d_{t+1}^-(\s_{t+1})\1_{\{\k_t^{\top}\r_{t+1}< 1\}}
\\
&+ d_{t+1}^+(\s_{t+1}) \1_{\{\k_t^{\top}\r_{t+1}\geq 1 \}}\big)+\nu \Big|\s_t\Big]\\
=&\max_{\nu>0} ~\Big\{-\frac{1}{4} \nu^2 d_t^-(\s_t) + \nu\Big\}.
\end{align*}
Note that the above problem is identical to the one in (\ref{eqd_t_minus}). Therefore, $\mathcal{D}_t(\bm \lambda,\nu)$ attains its maximum $\frac{1}{d_t^-(\s_t)}$ at $\bm \lambda^+ = -\nu^+ \k_t^-$ and $\nu^+=\frac{2}{d_t^-(\s_t)}$ where $\k_t^-$ is defined in problem (\ref{eqd_t_minus}).
If $\nu < 0$, identifying the optimal $\bm\lambda$ within the cone $-\bm \lambda \in\cA_t$ is equivalent to identifying the optimal $\k_t$ within the cone $\k_t\in\cA_t$ when we set $\bm \lambda =\nu \k_t $. Then, it has
\begin{align*}
&\max_{\nu<0, -\bm\lambda\in\cA_{t}}\mathcal{D}_t(\lambda,\nu) \\
=& \max_{\nu<0, \k_t \in\cA_{t}} \E\Big[-\frac{1}{4}\nu^2( 1+\k_t^{\top}\r_{t+1})^2 \big(d_{t+1}^-(\s_{t+1}) \1_{\{\k_t^{\top}\r_{t+1}\leq  -1\}} \\
&+ d_{t+1}^+(\s_{t+1}) \1_{\{\k_t^{\top} \r_{t+1}> -1 \}} \big) +\nu \Big|\s_t\Big]\\
=&\max_{\nu<0} \big\{-\frac{1}{4} \nu^2 \big\{\min_{\k_t \in\cA_{t}} \E\big[ ( 1+ \k_t^{\top} \r_{t+1})^2 \\
&\cdot\big( d_{t+1}^-(\s_{t+1}) \1_{\{\k_t^{\top}\r_{t+1}\leq  -1\}}+ d_{t+1}^+(\s_{t+1}) \1_{\{\k_t^{\top}\r_{t+1}> -1\}}\big)\\
&+\nu\Big|\s_t\big]\big\}\big\}.
\end{align*}
Then, we know that $\mathcal{D}_t(\lambda,\nu)$ attains its maximum $0$ when
$\nu$ approaches to $0$. Substituting $\bm \lambda^+=-\nu^+ \k_t^-$ and $\nu^+=\frac{2}{d_t^-(\s_t)}$ into
(\ref{eqnm}) yields the expression the optimizer for problem $(\cA_{\VSSM,t}^+)$ which is given by
\begin{align*}
&m_{t+1}^+=\frac{1}{d_{t}^-(\s_t)}\big( d_{t+1}^-(\s_{t+1}) (1-\r_{t+1}^{\top}\k_t^{-}) \1_{\{\r_{t+1}^{\top}\k_t^{-} \leq  1\}}\\
&~~~+ d_{t+1}^+(\s_{t+1}) (1-\r_{t+1}^{\top}\k_t^{-}) \1_{\{\r_{t+1}^{\top}\k_t^{-}>  1\}}\big).
\end{align*}
Using the above solution to problem gives the value function at time $t$ as $J_t(m_1,\ldots, m_t,\s_t) = \frac{1}{d_t^-(\s_t)} m_1^2  m_2^2 \dots m_t^2$. Applying a similar approach to for the case that $m_1 m_2 \cdots m_t<0$, we may obtain the solution 
$m_{t+1}^-$$=$$\frac{1}{d_{t}^+(\s_t)}\big[d{t+1}^-(\s_{t+1})(1+\r_{t+1}^{\top}\k_t^{+})\1_{\{\r_{t+1}^{\top}\k_t^{+} \geq  -1\}}
+d_{t+1}^+(\s_{t+1})(1+\r_{t+1}^{\top}\k_t^+) \1_{\{\r_{t+1}^{\top}\k_t^+< -1\}}\big]$, where $\k_t^+$ is defined in (\ref{eqd_t_plus}). Then, under this case, the value function is,
$J_t(m_1,\ldots, m_t,\s_t) = \frac{1}{d_t^+(\s_t)} m_1^2m_2^2\dots m_t^2$. Note that when $m_1m_2\cdots m_{t}= 0$, we set $m_{t+1}^*=m_{t+1}^+$. Thus, combining both of the above cases, we obtain the optimal solution as
\begin{align}
m_{t+1}^*&=m_{t+1}^+\1_{\{m_1m_2\cdots m_{t}\geq 
    0\}}+m_{t+1}^-\1_{\{m_1m_2\cdots m_{t}< 0\}}, \label{def_m_t+1}
\end{align}
with the value function being $J_t(m_1,\ldots, m_t,\s_t)$ $=$ $\frac{1}{d_t^-(\s_t)} m_1^2 \cdots m_{t}^2 \1_{\{m_1\cdots m_{t}\geq 0\}}$$
+$$\frac{1}{d_t^+(\s_t)}m_1^2\cdots m_{t}^2 \1_{\{m_1\cdots m_{t}< \ 0\}}$. We then proceed to prove that $m_1^* m_2^*\cdots m_{T}^*$ = $(d_0^-(\s_0))^{-1}\prod_{i=0}^{T-1}B_i(\s_i)$. Similarly, using the induction method, we prove that
\begin{align}
m_1^*m_2^*\cdots m_{t}^*= &(d_0^-(\s_0))^{-1}\prod_{i=0}^{t-1} B_i(\s_i) \big( d_t^-(\s_t) \1_{\{m_1^*\cdots
    m_{t}^*\geq  0\}}\notag\\
& +d_t^+(\s_t)\1_{\{m_1^*\cdots m_{t}^*<  0\}}\big), \label{VSSM_claim2}
\end{align}
which is the conditional expectation expression in (\ref{eqvssm_exp}). 
When $t=1$, the claim (\ref{VSSM_claim2}) is true, as it has $m_1^*=m_1^+=(d_0^-(\s_0))^{-1}  d_1^-(\s_1) (1-\r_1^{\top}\k_0^-(\s_0))
=(d_0^-(\s_0))^{-1}\prod_{i=0}^{0}B_i(\s_0)\big(d_1^-(\s_1)\big)$.
We {assume} that the claim (\ref{VSSM_claim2}) is true at time $t$ and prove that such claim also holds at time $t+1$. Using (\ref{def_m_t+1}), it has
\begin{align*}
&m_1^* \cdots m_{t+1}^* \\
=& (d_0^-(\s_0))^{-1} \prod_{i=0}^{t-1}B_i(\s_i) \\
&\cdot\big(d_t^-(\s_t) \1_{\{m_1^*\cdots
	m_{t}^*\geq 0\}}+d_t^+(\s_t) \1_{\{m_1^*\cdots m_{t}^*< 0\}}\big) \\
&\cdot\big(m_{t+1}^+\1_{\{m_1^*\cdots m_{t}^*\geq 0\}}+m_{t+1}^-\1_{\{m_1^*\cdots m_{t}^*< 0\}}\big)\\
=& (d_0^-(\s_0))^{-1}\prod_{i=0}^{t-1}B_i(\s_i)\big((1-\r_{t+1}^{\top}\k_t^-)\1_{\{m_1^*\cdots m_t^*\geq 0\}}\\
& +(1+\r_{t+1}^{\top}\k_t^+)\cdot \1_{\{m_1^*\cdots m_t^*< 0\}}\big) \\
& \cdot \big(d_{t+1}^-(\s_{t+1})\cdot \1_{\{m_1^*\cdots m_t^*m_{t+1}^*\geq 0\}}
+d_{t+1}^+(\s_{t+1})\\
&\cdot\1_{\{m_1^*\cdots m_t^*m_{t+1}^*<0\}}\big) \\
=&  (d_0^-(\s_0))^{-1}\prod_{i=0}^{t} B_i(\s_i) \big(d_{t+1}^-(\s_{t+1}) \1_{\{m_1^*\cdots m_t^*m_{t+1}^*\geq 0\}} \\
&+d_{t+1}^+(\s_{t+1})\1_{\{m_1^*\cdots m_t^*m_{t+1}^*<0\}}\big).
\end{align*}
Therefore, it has $m_1^*\cdots m_{T}^*$$=$$(d_0^-(\s_0))^{-1}\prod_{i=0}^{T-1}B_i(\s_i)$, which complete our proof of claim (\ref{VSSM_claim2}).\endproof

Theorem \ref{thm-VSSM} indicates that the density of VSSM given in (\ref{eq_VSMM_desity}) shares a similar product form as in markets with independent excess returns (see, \cite{cui2012better}). Our dynamic factor model complicates the expression of each term by introducing the possible FIO through $\k_t^-(\s_t)$ and $\k_t^+(\s_t)$, defined by problems (\ref{eqd_t_minus}) and (\ref{eqd_t_plus}), respectively.

With the help of Theorem \ref{thm-VSSM}, we obtain the following conditions under which the pre-committed optimal portfolio policy satisfies time consistency in efficiency.
\begin{theorem}\label{thmTCIE-VSSM} \sl
    When the VSSM of the market satisfies either one of the following conditions:
    \begin{align}
    ~\E \Big[\frac{\mathrm{d} \tilde{\mP}}{\mathrm{d} \mP}\Big|\mathcal{F}_t\Big]\geq 0, ~t=0,1,\dots,T, \label{eqcondition-1}
    \end{align}
    or
    \begin{align}
    ~\E \Big[\frac{\mathrm{d} \tilde{\mP}}{\mathrm{d} \mP} \Big|\mathcal{F}_k\Big]=\E \Big[\frac{\mathrm{d} \tilde{\mP}}{\mathrm{d} \mP}\Big|\mathcal{F}_{\tau}\Big]< 0,~k=\tau,\dots,T, \label{eqcondition-2}
    \end{align}
    where the stopping time $\tau$ is defined as
    $\tau$$=$$\inf\big\{t \Big\vert ~\E \big[\frac{\mathrm{d} \tilde{\mP}}{\mathrm{d} \mP} \Big| \mathcal{F}_t \big]< 0,~t= 1,\ldots,T\big\}$, the pre-committed optimal portfolio policies (\ref{eqopt_policy_P1}) and (\ref{eqopt_policy_P2}) satisfy Time Consistency in Efficiency. 
\end{theorem}  

\proof We first rewrite the policy (\ref{eqopt_policy_P2}) in more concise for by using the notation $\lambda^*$ in (\ref{eqlambda_opt}) as follows
\begin{align}
&\bm{\pi}^{*}(x_t, \s_t) = \big(r_{t+1}^0x_t - \frac{x_{\tg} +\lambda^* }{\rho_{t+1}}\big)
\Big( -\k_t^-(\s_t) \1_{\{\rho_tx_t \leq x_{\tg} + \lambda^* \}}\notag\\
&+\k_t^+(\s_t) \1_{\{\rho_tx_t \leq x_{\tg} + \lambda^* \}}\Big)\label{eqopt_policy_P2_lambda}
\end{align}
for $t=0,\ldots, T-1$. Note that, using the expression in (\ref{eqvssm_exp}), we know that condition (\ref{eqcondition-1}) is equivalent to 
$\r_{t+1}^{\top}\k_t^-(\s_t) \leq 1$ for $t=0,1,\dots,T-1$. We then prove that the wealth process achieved by the efficient portfolio policy in (\ref{eqopt_policy_P2}) satisfies 
\begin{align}
x_{\tg} +\lambda^* \geq \rho_t x_t, ~~\forall~ t=0,1,\dots,T-1. \label{thmTCIE_claim} 
\end{align}
We prove the above result by induction method. At time $t=0$, due to our assumption $x_{\tg}> \rho_0 x_0$ and the fact $\lambda^*>0$, it has $x_{\tg} + \lambda^* \geq \rho_0 x_0$. At time $t$, suppose it has  $ x_{\tg} + \lambda^* \geq \rho_t x_t$ holds and we prove the claim (\ref{thmTCIE_claim}) is also true at $t+1$. Applying the optimal policy (\ref{eqopt_policy_P2_lambda}), at time $t+1$, we have
$\rho_{t+1} x_{t+1}-(x_{\tg} + \lambda^*)$$=$$\rho_{t+1} r_{t+1}^0 x_t -( x_{\tg} + \lambda^* )
+ \rho_{t+1} \r_{t+1}^{\top} (-\k_t^-(\s_t))(r_{t+1}^0 x_t - \rho_{t+1}^{-1}(x_{\tg}+\lambda^*)) $$=$$(1 -  \r_{t+1}' \k_t^-(\s_t))(\rho_t x_t  - (b-\lambda^-))\leq 0$,
\normalsize
which implies that $(x_{\tg} + \lambda^*) \geq \rho_{t+1}x_{t+1}$.

On the other hand, we consider the following truncated version of problem $\cP_{\mv}^2(x_{tg})$ starting from time $k\in \{1,\ldots, T-1\}$ with give wealth level $x_k>0$ and target wealth level $b_k>0$,
\begin{align*}
\cP(b_k, x_k)|_{k}^{T} ~ \min_{\bm \pi_k,\dots,\bm \pi_{T-1}} &~ \Var_k(x_T)\\
\mbox{s.t.}&~\E[x_T|\cF_k] = b_k ,\\ 
&~~~\textrm{$\{x_t, \bm{\pi}_t,\r_t,\s_t \}$ {satisfies} (\ref{eqr}),(\ref{eqs}),(\ref{def_xt})}
\end{align*}
Using Theorem \ref{thm_policy_P2}, we obtain the optimal policy for problem $\cP(b_k, x_k)|_{k}^{T}$ as 
$\bm{\hat{\pi}}_t(x_t,\s_t) $ $=$ $\big[-\k_t^-(\s_t)\1_{\{\rho_t x_t\leq  b_k + \lambda_k \}} 
                                   +\k_t^+(\s_t) \1_{\{\rho_t x_t> b_k +\lambda_k\}}\big] 
                                   \big( r_{t+1}^0 x_t - \rho_{t+1}^{-1} (b_k +\lambda_k)\big).$
When $b_k \geq \rho_k x_k$, the optimal portfolio policy of original problem $\cP_{\mv}^2(x_{tg})$ is also efficient for $\cP(b_k, x_k)|_{k}^{T}$ with $\lambda_k = \frac{\rho_k x_{k}-b_k}{1-(d_k^-(\s_k))^{-1}}$. We further have $b_k \geq \rho_k x_{k}$$~\Leftrightarrow~$$(b_k-\rho_k x_{k})\frac{1}{1- b_k^-(\s_k)}\geq 0 $ $~\Leftrightarrow~$ $b_k-\frac{b_k-\rho_k x_{k}}{1-(b_k^-(\s_k))^{-1}}\geq \rho_k x_k~$ $\Leftrightarrow$ $~b_k-\lambda_k \geq \rho_ k x_k$.
Therefore, to examine whether the truncated optimal policy (\ref{eqopt_policy_P2}) is efficient for $\cP(b_k,x_k)|_{k}^T$, we can just set $x_{\tg} + \lambda^* = b_k +\lambda_k $ and check whether it has $b_k + \lambda_k \geq \rho_k x_k$ or not. Obviously, under condition (\ref{eqcondition-1}), we have $x_{\tg} + \lambda^*\geq \rho_{t+1} x_{t+1}$, which implies $b_k + \lambda_k \geq \rho_k x_k$. In other words, under condition (\ref{eqcondition-1}), the truncated optimal policy (\ref{eqopt_policy_P2}) is efficient for the truncated problem $\cP(b_k,x_k)|_{k}^T$. We then consider condition (\ref{eqcondition-2}), which is equivalent to three conditions: (i) $\r_{t+1}^{\top}\k_t^-(\s_t) \leq 1$ for $t=0,1,\dots,\tau-1$; (ii)$\r_{t+1}^{\top}\k_t^-(\s_t) > 1$ for $t=\tau$; and (iii) $\r_{t+1}^{\top}\k_t^+(\s_t) = 0$ for $t=\tau+1,\dots, T-1$. We then check each of these cases. Based on the analysis for assumption (\ref{eqcondition-1}), condition (i) implies that $(x_{\tg} +\lambda^*) \geq \rho_t x_t$ for $t=0,1,\dots,\tau$, which further ensures that the truncated optimal policy (\ref{eqopt_policy_P2}) is efficient for  $\cP(b_t,x_t)|_{t}^T$, $t=0,1,\dots,\tau$. Conditions (ii) implies that 
$\rho_{\tau+1} x_{\tau+1} - (x_{\tg} + \lambda^*)$$=$$\rho_{\tau+1} r_{\tau+1}^0 x_\tau -(x_{\tg} + \lambda^*)+  \rho_{\tau+1} \r_{\tau+1}^{\top} (-\k_\tau^-(\s_\tau)) (r_{\tau+1}^0 x_\tau - \rho_{\tau+1}^{-1}(x_{\tg}+\lambda^*))
$$=$$(1 -  \r_{\tau+1}^{\top} \k_\tau^-(\s_\tau))(\rho_\tau x_\tau  - ( x_{\tg} + \lambda^*))> 0$.

Similar to the above formula, condition (iii) implies that $\rho_{t+1}x_{t+1}-( x_{\tg} + \lambda^* )$$=$$\rho_t x_t  - (x_{\tg} + \lambda^*) > 0$, for $t=\tau+1,\dots, T-1$, and $\k_t^+(\s_t) = \0$ for $t=\tau+1,\dots, T-1$. According to the optimal policy in (\ref{eqopt_policy_P2}), we have  $\bm \pi_{t}^{*}(x_t,\s_t) = \0$ for $t=\tau+1,\dots,T-1$ which is the minimum variance policy as well as efficient for problem $\cP_t(b_t,x_t)|_{t}^T$ for $t=\tau+1,\dots,T-1$.\endproof

Theorem \ref{thmTCIE-VSSM} extends the related result in \cite{cui2017mean} to the dynamic factor model-based market. At first glance, conditions (\ref{eqcondition-1}) and (\ref{eqcondition-2}) appear quite similar to those in \cite{cui2017mean}. However, they differ because the density defined in (\ref{eq_VSMM_desity}) is characterized by the parameters $d_0^-(\s_0)$, {$\{\k_t^-(\s_t)\}|_{t=0}^{T}$, and $\{\k_t^+(\s_t)\}|_{t=0}^T$} which are further related to (\ref{eqs}). In other words, the choice of the dynamic factor model in (\ref{eqs}) may lead to different versions of the conditions (\ref{eqcondition-1}) and (\ref{eqcondition-2}).

\section{Empirical Analysis}\label{sec_empirical_analysis}

In this section, we apply the portfolio policy developed in Section \ref{sec_mv_policy} to the constrained MMV portfolio optimization problem, incorporating Fama-French's five factors: market (MKT), size (SMB), value (HML), investment (CMA), profitability (RMW), along with the momentum (MOM) factor (see \cite{fama2015five}). We analyze ten industrial portfolios: NoDur (N), Durbl (D), Manuf (M), Enrgy (E), HiTec (Hi), Telcm (T), Shops (S), Hlth (Hl), Utils (U), and Other (O). The return model (\ref{eqr}) is specified as $\r_{t} = \bm{\alpha} + \bm{B}\bm{s}_{t} + \bm{\epsilon}_{t}$, where $\r_t = \big( r^{N}_t, r^{D}_t, r^{M}_t, r^{E}_t, r^{Hi}_t, r^{T}_t, r^{S}_t, r^{Hl}_t, r^{U}_t, r^{O}_t \big)^{\top}$ represents the excess returns of ten industry portfolios,
$\s_{t} = \big( \textrm{MKT}_t, \textrm{SMB}_t, \textrm{HML}_t, \textrm{CMA}_t, \textrm{RMW}_t, \textrm{MOM}_t \big)^{\top}$
is the state vector constructed by six factors, $\bm B\in \mR^{10\times 6}$ represents the factor loading matrix, and $\bm{\epsilon}_{t}\in \mR^{10}$ denotes the noise with a zero mean and a covariance matrix $\bm{\Sigma}_{\epsilon}$.  According to the dynamic factor model (see \cite{stoyanov2021dynamics,zhao2019revisiting}), the state equation (\ref{eqs}) is specified as $\s_t = \big(\bm{I} - \bm{\Phi}\big)\s_{t-1} + \bm{\xi}_t$ for $t=0,\ldots,T-1$, where $\bm{I}$ is the identity matrix, $\bm{\Phi}$ is the matrix of mean-reversion coefficients, and $\bm{\xi}_t \in \mR^6$ is the noise term with zero mean and covariance matrix $\bm{\Sigma}_{\bm{\xi}}$. Using monthly data on returns and factors covering the period from July 1963 to May 2024, we employ a regression method to estimate all parameters in the return model and factor model.\footnote{The historical data of industrial portfolio and factors are from Kenneth French's website: \url{http://mba.tuck.dartmouth.edu/pages/faculty/ken.french/data\_library.html}. Using this data, we calibrate the dynamic factor model, where the coefficients are given at {\url{https://github.com/JinChengneng/Dynamic-Factor-MV}}.}

We solve problem $\cP^2_{\mv}(x_{\tg})$ with $T=6$ by utilizing Algorithm \ref{alg-factor}. To implement Algorithm \ref{alg-factor}, we use the historical data of state variables (731 samples), and simulate $L=1000$ samples of $(\s_t,\r_t)$ for each of the state's realization. In this algorithm, one key step is to characterize the functions $d_{t}^-(\s_t)$ and $d_{t}^+(\s_t)$ from discrete data points for each $t$. We use the neural network to fit these functions where the architecture comprises three hidden layers with dimensions of $8$, $16$ and $8$, respectively. The activation functions employed are the Rectified Linear Unit (ReLU) for both hidden layers, enhancing non-linearity, and a sigmoid activation function for the output layer to ensure $d_t^-(\s_t),d_t^+(\s_t)\in[0,1]$ (see Lemma \ref{lem-inequ}). In our numerical experiment, the network exhibits robust performance, achieving validation errors below $1 \times 10^{-4}$, indicating a high degree of accuracy in approximations post-training.

\begin{figure}
\small
    \centering
    \begin{subfigure}[b]{0.4\textwidth}
        \includegraphics[width=0.85\textwidth]{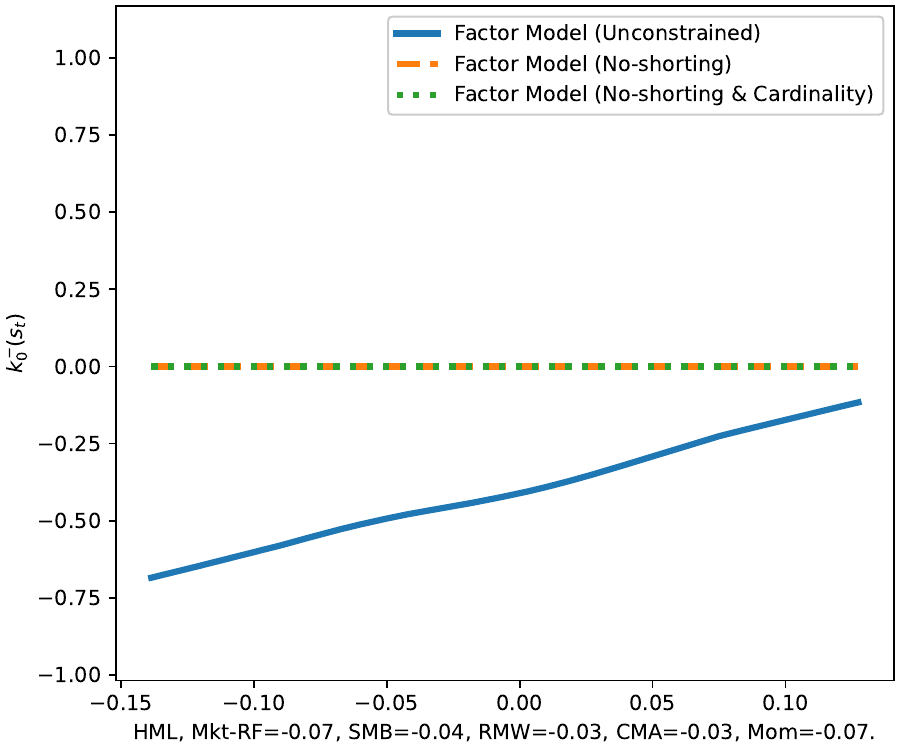}
        \caption{\footnotesize Case 1.}
    \end{subfigure}
    \begin{subfigure}[b]{0.4\textwidth}
        \includegraphics[width=0.85\textwidth]{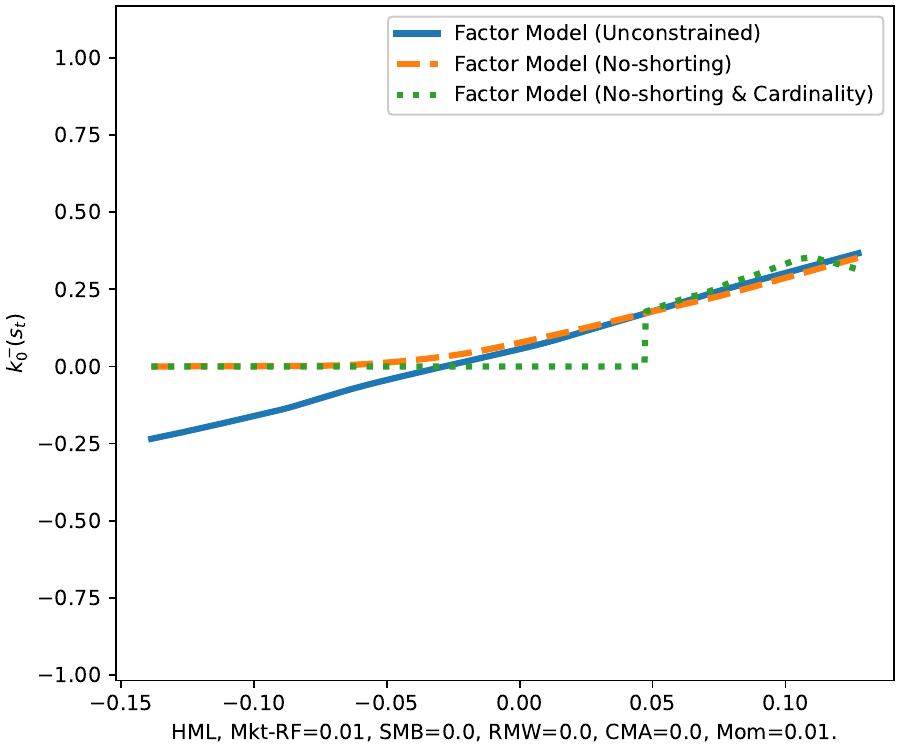}
        \caption{\footnotesize Case 2.}
    \end{subfigure}
    \begin{subfigure}[b]{0.40\textwidth}
        \includegraphics[width=0.85\textwidth]{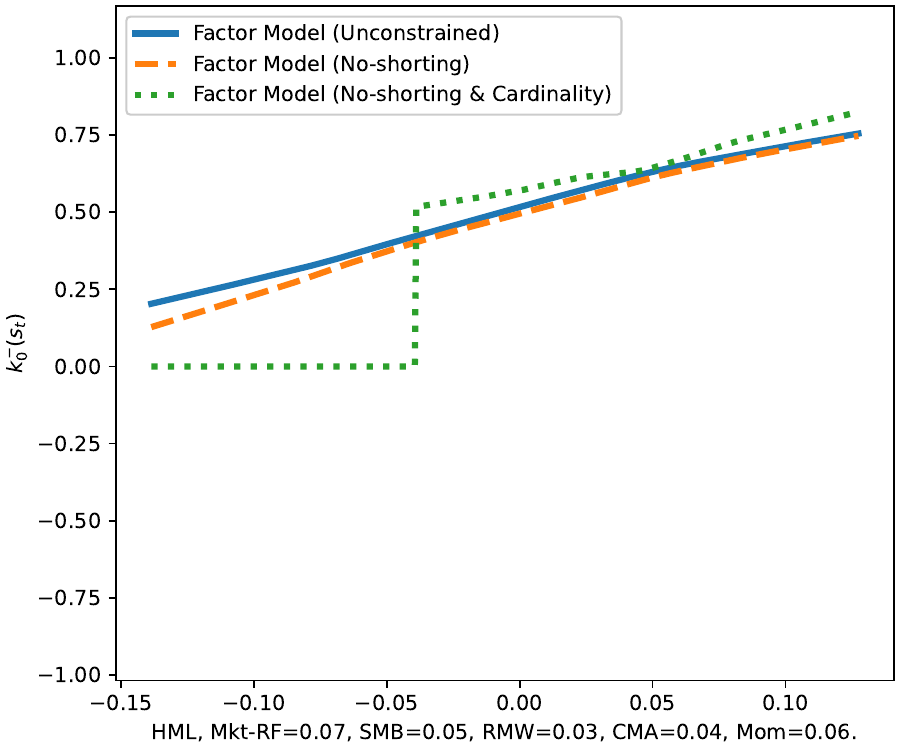}
        \caption{\footnotesize Case 3.}
    \end{subfigure}
    \caption{\small Portfolio policy coefficients $\k_0^-$ of Energy (E) industry with HML factor.}
    \label{fig:k_HML_asset_3}
\end{figure}

After fitting the FIO functions ${d_t^-(\s_t)}|_{t=0}^{T}$ and ${d_t^+(\s_t)}|_{t=0}^T$, we can compute the portfolio allocation vector $\{\k_t^-(\s_t)\}$, $\{\k_t^+(\s_t)\}$ as function of state $\s_t$. To better express the impact of the  market state $\s_t$ on the portfolio allocation, we plot the entry Enrgy (E) in $\k_t^-$ in response to the state HML in Figure \ref{fig:k_HML_asset_3}. {Specifically}, we vary state HML and fix all other five state variables in three different cases, i.e.,  Case 1, Case 2 and Case 3 {represent} a scenario where the five inactive state variables are at the 5th percentile, median value and  95th percentile in historical data, respectively. In Figure \ref{fig:k_HML_asset_3}, we compare three different {portfolios}, namely, the unconstrained portfolio, portfolio with no-shorting constraints, and the portfolio with {cardinality} and no-shorting constraints. We can see that the factor HML (High Minus Low) and the sector Energy industry exhibit a positive relationship.  The most striking feature is the stark difference between the constrained (dashed line and dotted line) and unconstrained (solid line) policies. The no-shorting constrained policy closely follows the unconstrained policy when allocations are positive in most cases but flattens at zero, preventing negative allocations. The cardinality-constrained policy exhibits a step function behavior, abruptly switching from zero allocation to large (see Figure \ref{fig:k_HML_asset_3}) at a certain threshold. This behavior is due to the cardinality constraint, which forces the portfolio to make discrete decisions about including or excluding the specific sector.

\begin{figure}
    \centering
    \begin{subfigure}[b]{0.46\textwidth}
        \begin{subfigure}[b]{\textwidth}
            \centering
            \includegraphics[width=0.85\textwidth]{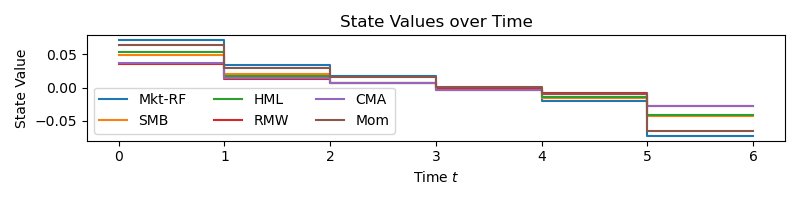}
            \includegraphics[width=0.85\textwidth]{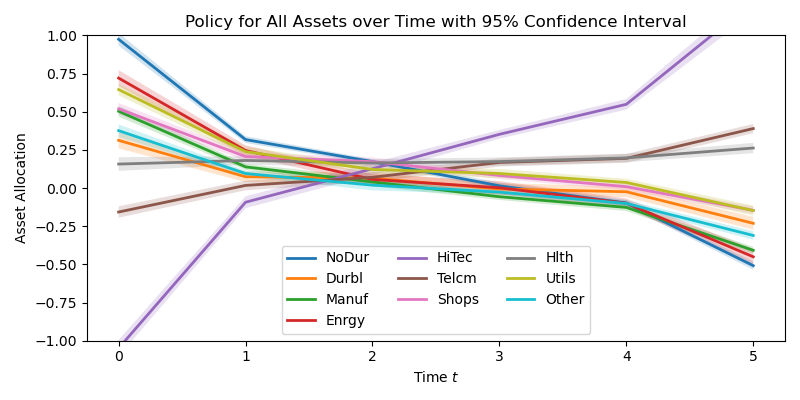}
            \caption{\footnotesize States and Policies.}
            \label{fig_policy_traj}
        \end{subfigure}
    \end{subfigure}
    \begin{subfigure}[b]{0.46\textwidth}
        \centering
        \includegraphics[height=170pt, width=0.85\textwidth]{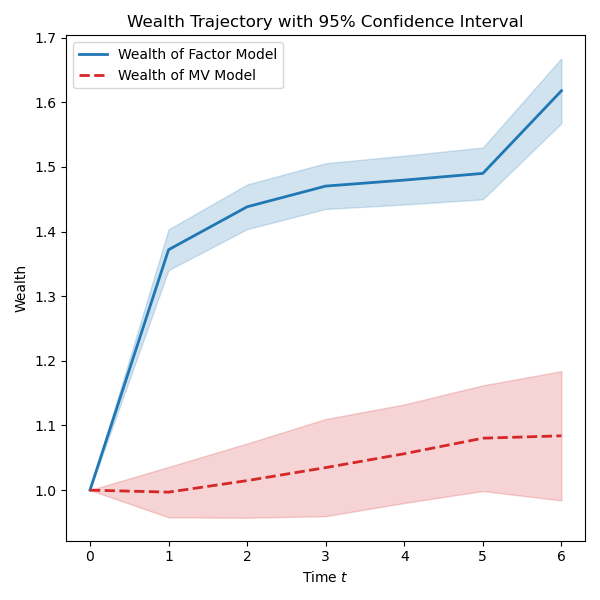}
        \caption{\footnotesize Wealth.}
        \label{fig_wealth_traj}
    \end{subfigure}
    \caption{\small Trajectories of States, Actions and Wealth of Factor Model and Traditional MV Model.}
    \label{fig_traj}
\end{figure}

Figure \ref{fig_traj} further illustrates the trajectories of states, portfolio {decision}, and wealth for our {factor-based} MMV model.\footnote{To simulate deteriorating market conditions, we select state values corresponding to the 95th, 75th, 60th, 40th, 25th, and 5th percentiles of each factor as time progresses, as depicted in Figure \ref{fig_policy_traj}.} This approach allows us to evaluate the models' performance under increasingly challenging market scenarios. Figure \ref{fig_policy_traj} demonstrates how our policy responds dynamically to these shifting market states. Notably, the portfolio weights for most assets, such as Non-Durables (NoDur) and Energy (Enrgy), exhibit a downward trend as market conditions worsen. This behavior can be attributed to the increased risk and potentially lower expected returns associated with these sectors during market downturns. Conversely, we observe an increase in the allocation to certain assets, particularly High Technology (HiTec). This shift likely reflects the model's assessment that technology stocks may offer better risk-adjusted returns or serve as a hedge against adverse market movements in the simulated scenario.

The divergent trajectories of asset allocations underscore our model's capacity to adapt to changing market factors, resulting in a more nuanced and potentially robust portfolio construction. The efficacy of this dynamic approach is further evidenced in Figure \ref{fig_wealth_traj}, which displays the wealth trajectories for our model and the MMV model ignoring the factor structure. This  MMV model ignoring the factor structure is just the model adopted in \cite{li2000optimal}, which is called as \textit{benchmark MV model} in the following part. Our model exhibits superior performance in two key aspects: higher profitability (larger mean wealth trajectory) and greater robustness (smaller confidence interval) compared to the traditional MV model.

\begin{figure}[h]
    \centering
    \includegraphics[width=0.4\textwidth]{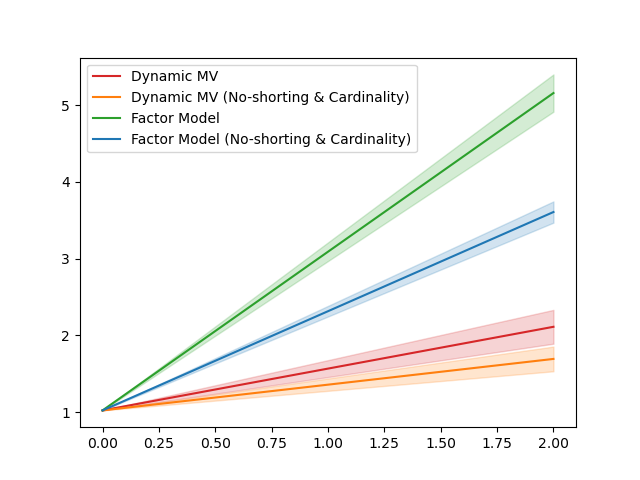}
    \caption{\small Efficient frontiers of varying models with/without cone constraints.}
    \label{fig:eff_frt}
\end{figure}

Using the real market data, we compare the out-of-sample performance of our factor model-based policy with the policy resulted from the benchmark MV model. Figure \ref{fig:eff_frt} reports the out-of-sample MV efficient frontier of different models. Additional performance metrics of portfolios (including, mean value ({Mean}), standard {deviation} (STD), {Sharpe Ratio (Sharpe)}, {Sortino} Ratio {(STR)}, {Conditional-value-at-Risk (CVaR)} and Value-at-Risk (VaR)) of unconstrained model (UC), no-shorting model (NS) and no-shorting \& cardinality model (N\&C) are shown in Table \ref{tbl-performance}. All these statistics, further demonstrate that almost all metrics favor our dynamic factor-based MMV portfolio model over the benchmark MV model when subject to the same constraints.

\begin{table}[h]
\centering
\caption{\small Performance comparision between different models}\label{tbl-performance}
\footnotesize
\begin{tabular}{c ccc ccc}
\toprule
& \multicolumn{3}{c}{IID returns} & \multicolumn{3}{c}{Factor model}\\
               \cmidrule(lr){2-4}     \cmidrule(lr){5-7}  
Criteria &   UC    & NS &  N\&C & UC    & NS & N\&C  \\
\cmidrule(lr){1-1} \cmidrule(lr){2-4}     \cmidrule(lr){5-7}  
Mean     &  1.218  &  1.107   &  1.107         &  1.419  &  1.300    & 1.282 \\
STD      &  0.424  &  0.293   &  0.293         &  0.202  &  0.227    & 0.216 \\
Sharpe   &  0.462  &  0.291   &  0.291         &  1.968  &  1.226    & 1.207\\
STR      &  0.22   &  0.096   &  0.096         &  0.311  &  0.963    & 0.949\\
CVaR     & -0.026  & -0.035  & -0.035         & -1.061  & -0.897    & -0.893\\
VaR      & -0.345  & -0.594   & -0.594         & -1.121  & -0.963    & -0.949\\
\bottomrule
\end{tabular}
\label{tbl:performance}
\end{table}

\begin{figure}
    \centering
    \includegraphics[width=0.7\linewidth]{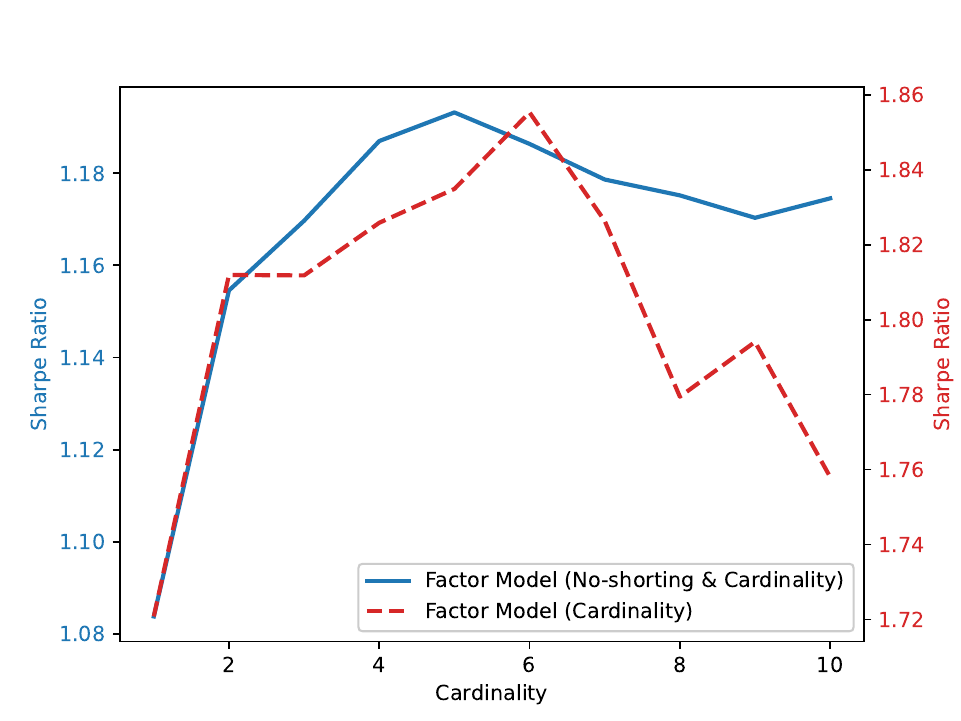}
    \caption{\small Sharpe Ratios of varying cardinality $q$ with management fee and transition cost.}
    \label{fig:vary-q}
\end{figure}

At last, we examine the impact of the cardinality {parameter} $q$ on the performance of different portfolio policies. In practical market scenarios, two key factors significantly impact the performance of dynamic portfolio models: management fees and transition costs.\footnote{Portfolio managers or financial advisors typically levy management fees, which remain constant relative to the investment timeline. Conversely, transition costs fluctuate in proportion to the volume of portfolio adjustments and include expenses like brokerage commissions and taxation associated with rebalancing or modifying a {portfolio. Both} these expense categories are intrinsically linked to the cardinality parameter $q$.} To quantify the effect of these expenses, we adopt similar model in \cite{gao2015time}. The management fee is represented by $M(x_0, q)=\alpha_0 \cdot q \cdot x_0$, where $\alpha_0$ is the unit fee. For transaction costs at time $t$, we calculate it based on changes in asset holdings: {$TC_t = \alpha_1 \cdot \sum_{i=1}^N \big| \bm{\pi}_{t}^i(x_t, s_t)/S_{t}^{i} - \pi_{t-1}^{i}(x_t, s_t)/S_{t-1}^{i}\big|$}, where $\alpha_1$ is the unit cost. In our analysis, we set $\alpha_0=0.2\%$ and $\alpha_1=0.02\%$. Figure \ref{fig:vary-q} illustrates the changes of the Sharpe Ratio when accounting for these costs. The graph reveals that a low cardinality results in inadequate portfolio diversification. Conversely, high cardinality introduces excessive assets, leading to higher {management} cost. Thus, identifying the optimal cardinality parameter $q$ is crucial for maximizing the dynamic MV portfolio's profitability.

\section{Conclusion}\label{sec_conclusion}

This paper examines the MMV portfolio selection problem in a dynamic factor model-based market with cone constraints. The returns of risky assets are driven by a vector of state variables, encompassing models like the linear factor model, Markov regime-switching, GARCH, and GARCH-jump mixture models as special cases. We derive a semi-analytical pre-committed optimal portfolio policy, showing that the piecewise linear feedback policy is highly dependent on the stochastic FIO process. Additionally, we provide an economic interpretation and propose various numerical algorithms to compute the portfolio policy under different market conditions. Moreover, we derive the variance-optimal signed supermartingale measure (VSSM) and establish conditions under which the pre-committed optimal portfolio policy is time-consistent in efficiency (TCIE). Our empirical analysis using US market data for the factor model not only validates the theoretical results and numerical algorithms but also demonstrates the practical value of the proposed framework.

Our framework has some limitations, as it assumes that the state variables (factors or market states) are observable. It can be extended to handle settings with latent variables, such as those in stochastic volatility models, using filters or observers developed for control problems with incomplete information. Additionally, while this paper focuses on the pre-committed policy, exploring the time-consistent policy within this framework is another important avenue for future research.

\section*{References}
\bibliographystyle{IEEEtran}
\bibliography{mv_factor}

\end{document}